\documentclass[sigconf]{acmart}
\copyrightyear{2026}
\acmYear{2026}
\setcopyright{cc}
\setcctype{by}
\acmConference[SoCC '26 Vol. 2]{ACM Symposium on Cloud Computing Volume 2}{November 18--20, 2026}{Singapore, Singapore}
\acmBooktitle{ACM Symposium on Cloud Computing Volume 2 (SoCC '26 Vol. 2), November 18--20, 2026, Singapore, Singapore}
\acmDOI{10.1145/3828161.3857885}
\acmISBN{979-8-4007-2816-7/2026/11}
\AtBeginDocument{%
  }
\usepackage[]{hyperref}
\usepackage{subcaption}
\usepackage{textcomp}
\usepackage{xspace}
\usepackage{enumitem}
\usepackage{fancyhdr}
\usepackage{xcolor}
\definecolor{whitesmoke}{RGB}{230, 230,230}
\newcommand{\reqbox}[1]{\par\smallskip\noindent{\setlength{\fboxsep}{1mm}\colorbox{whitesmoke}{\parbox{\dimexpr\linewidth-2\fboxsep\relax}{#1}}}\par\smallskip}
\usepackage{algorithm}
\usepackage{algorithmic}
\usepackage{pifont}
\usepackage{multirow} 
\usepackage{makecell}
\newcommand{\sys}{TokenScale\xspace}
\newcommand{\aibrix}{AIBrix\xspace}
\newcommand{\blitz}{BlitzScale\xspace}
\newcommand{\dist}{DistServe\xspace}
\newcommand{\spw}{SplitWise\xspace}
\newcommand{\sllm}{ServerlessLLM\xspace}

\newcommand{\llama}{Llama-3.1-8B\xspace}
\newcommand{\Qwen}{Qwen-3-30B-A3B\xspace}

\ifdefined\RELEASE
  \newcommand{\ignore}[1]{}
  \newcommand{\fixme}[1]{}
  \newcommand{\rui}[1]{}
  \newcommand{\siy}[1]{}
  \newcommand{\dmi}[1]{}
  \newcommand{\yix}[1]{}
  \newcommand{\yao}[1]{}
  \newcommand{\luo}[1]{}
  \newcommand{\leo}[1]{}
  \newcommand{\cz}[1]{}
  \newcommand{\hong}[1]{}
  \newcommand{\colin}[1]{}
  \newcommand{\TODO}[1]{}

\else
  \newcommand{\ignore}[1]{}
  \newcommand{\fixme}[1]{{\textcolor{red}{[~FIXME:~#1~]}}}
  \newcommand{\rui}[1]{{\textcolor{brown}{[~R:~#1~]}}}
  \newcommand{\siy}[1]{{\textcolor{purple}{[~S:~#1~]}}}
  \newcommand{\dmi}[1]{{\textcolor{blue}{[~D:~#1~]}}}
  \newcommand{\colin}[1]{{\textcolor{orange}{[~CL:~#1~]}}}
  \newcommand{\hong}[1]{{\textcolor{teal}{[~HR:~#1~]}}}
  \newcommand{\luo}[1]{{\textcolor{magenta}{[~LU:~#1~]}}}
  \newcommand{\leo}[1]{{\textcolor{green}{[~LE:~#1~]}}}
  \newcommand{\cz}[1]{{\textcolor{violet}{[~CZ:~#1~]}}}
  
  \newcommand{\TODO}[1]{{\textcolor{red}{TODO:~#1}}}
  
\fi
\begin{document}
\sloppy
% Ensure letter paper
% \pdfpagewidth=8.5in
% \pdfpageheight=11in

%%%%%%%%%%%---SETME-----%%%%%%%%%%%%%
% \newcommand{\sospsubmissionnumber}{8}
%%%%%%%%%%%%%%%%%%%%%%%%%%%%%%%%%%%%

\pagenumbering{arabic}
% \title{TokenPipe: Rapid and Fine-grained Autoscaling for Disaggregated LLM Serving}
% \title{Accurate and Timely Scaling Prefill/Decode Disaggregated LLM Serving Systems with Token Velocity and Convertible Decoders}
% \title{Autoscaling Disaggregated LLM Serving via\\ Token Velocity and Convertible Decoders}
% \title{\sys: Timely and Accurate Autoscaling for Disaggregated LLM Serving with Token Velocity}
% \title{\sys: Stage-aware Autoscaling for Disaggregated LLM Serving with Token Velocity}
\title{\sys: Timely and Accurate Autoscaling for Disaggregated LLM Serving with Token Velocity}
% \subtitle{\sout{Research Full}}

% [Camera-ready] Author list follows the HotCRP submission order.
\settopmatter{authorsperrow=5}
\author{Ruiqi Lai}
\authornote{Both authors contributed equally to this work.}
\affiliation{
  \institution{NTU Singapore}
  \country{Singapore}
}
\author{Hongrui Liu}
\authornotemark[1]
\affiliation{
  \institution{NTU Singapore}
  \country{Singapore}
}
\author{Chengzhi Lu}
\authornote{Corresponding author.}
\affiliation{
  \institution{NTU Singapore}
  \country{Singapore}
}
\author{Siyu Cao}
\authornote{Work done while at NTU Singapore.}
\affiliation{
  \institution{SMU}
  \country{Singapore}
}
\author{Li Leqi}
\authornotemark[3]
\affiliation{
  \institution{Huawei}
  \country{China}
}
\author{Zonghao Liu}
\affiliation{
  \institution{NTU Singapore}
  \country{Singapore}
}
\author{Siyang Shao}
\authornotemark[3]
\affiliation{
  \institution{Georgia Tech}
  \country{USA}
}
\author{Yixin Zhang}
\authornotemark[3]
\affiliation{
  \institution{C.F. Investment}
  \country{China}
}
\author{Luo Mai}
\affiliation{
  \institution{Univ. of Edinburgh}
  \country{United Kingdom}
}
\author{Dmitrii Ustiugov}
\affiliation{
  \institution{NTU Singapore}
  \country{Singapore}
}
\renewcommand{\shortauthors}{Lai et al.}
\renewcommand{\authors}{Ruiqi Lai, Hongrui Liu, Chengzhi Lu, Siyu Cao, Li Leqi, Zonghao Liu, Siyang Shao, Yixin Zhang, Luo Mai, and Dmitrii Ustiugov}

% \author{\normalsize{SOSP 2026 Submission
    % \textbf{\#\sospsubmissionnumber} -- Confidential Draft -- Do NOT Distribute!!}}
% \author{
% {\rm Ruiqi Lai$^{1,\dagger}$\quad Hongrui Liu$^{1,\dagger}$\quad Chengzhi Lu$^1$\quad Zonghao Liu$^1$\quad Siyu Cao$^{1}$\quad Siyang Shao$^{3,*}$\quad Yixin Zhang$^{4,*}$
% \\ \quad Luo Mai$^2$\quad Dmitrii Ustiugov$^1$\quad} 
% \\
% \textit{$^1$NTU Singapore \quad $^2$University of Edinburgh \qual $^3$Georgia Institute of Technology \qual $^4$Alibaba Group}
% \\
% \textit{$^*$Equal contribution}
% }
% \settopmatter{printfolios=true}
\begin{abstract}

The architectural shift to prefill-decode disaggregation in LLM serving improves resource utilization but struggles with the bursty nature of modern workloads. Existing autoscaling policies, often retrofitted from monolithic systems like those in AIBrix and DistServe, rely on lagging indicators such as GPU utilization or coarse-grained request counts. This results in slow reactions to load spikes, leading to significant SLO violations and costly overprovisioning. We introduce \emph{\sys}, an autoscaling framework that resolves this performance mismatch through two innovations. First, we propose \emph{Token Velocity}, a novel metric that unifies the prefill, network, and decode stages by quantifying their rate of work. As a leading indicator of system backpressure, it enables proactive scaling. Second, \emph{Convertible Decoders} allow decoder GPUs to dynamically execute prefill tasks during traffic spikes, creating a rapid-response buffer that absorbs bursts and eliminates the initialization latency of new prefillers. By uniting a predictive metric with a flexible system design, \sys significantly boosts the performance and efficiency of disaggregated LLM serving infrastructure. Our evaluation with production traces demonstrates that \sys improves the performance/cost by $1.3$-$6.4\times$ over state-of-the-art systems, including DistServe, BlitzScale, AIBrix, and SplitWise.
\end{abstract}



% \keywords{LLM serving, prefill-decode disaggregation, autoscaling, bursty workloads, token velocity}

\maketitle % should come after the abstract
% \def\thefootnote{$\dagger$}\footnotetext{These authors contributed equally to this work.}\def\thefootnote{\arabic{footnote}}\\
% \def\thefootnote{$*$}\footnotetext{Completed when these authors were at NTU Singapore.}\def\thefootnote{\arabic{footnote}}
% \pagestyle{plain} % should come right after \maketitle
% \author{
% {\rm Ruiqi Lai$^1$\quad Siyang Shao$^1$\quad Yixin Zhang$^1$\quad Fu Yao$^2$\quad Luo Mai$^2$\quad Dmitrii Ustiugov$^1$\quad} 
% \\
% \textit{$^1$NTU Singapore \quad $^2$University of Edinburgh}
% }

% \pagestyle{firstpage} % should come right after \maketitle

% \pagestyle{plain} % should come right after \maketitle

% \thispagestyle{firstpage}
\section{Introduction}

% \dmi{lacks an entry sentence on how dominant LLM inference is as a workload}
Large language model (LLM) inference has rapidly emerged as a dominant workload in modern AI infrastructure, placing substantial pressure on GPU efficiency, memory capacity, and request scheduling in production serving systems. 
The architectural shift to prefill/decode (PD) disaggregation is reshaping 
% Large Language Model~(LLM)
LLM serving, promising to optimize expensive accelerator resources by separating the compute-intensive prefill and memory-intensive decode stages~\cite{patel:splitwise,zhong:distserve,qin:mooncake}. This separation is designed to prevent performance interference and enable fine-grained scaling based on the distinct resource needs of each phase~\cite{zhong:distserve,zhang:blitzscale}.

Serving LLM inference efficiently at scale is a challenging problem: modern LLM workloads are often bursty, characterized by short-term albeit highly frequent spikes in requests as well as input and output tokens~\cite{choukse:dynamollm,wang2025burstgpt}. 
Azure production trace~\cite{choukse:dynamollm} reveals that a system can experience traffic bursts during 47\% of its operational time, with each burst lasting only 2.3 seconds on average.
This creates a severe performance mismatch, as policies designed to manage disaggregated resources are often ill-equipped for such dynamic traffic. The consequence is not just poor performance, but significant Service Level Objective~(SLO) violations in Time-to-First-Token (TTFT) and Time-Per-Output-Token (TPOT), often forcing operators into substantially overprovisioning of their expensive GPU clusters. 
An effective scaling policy in a large-scale LLM serving must react to traffic changes both accurately and rapidly. 
In a PD-disaggregated system, this requires rapidly scaling upstream prefillers to handle resource bursts, while equipping downstream decoders with accurate leading indicators.

% \dmi{This para is too long, see my comment below for a fix}
Unfortunately, existing strategies fail on both fronts because they are fundamentally mismatched with the PD disaggregation architecture.
For scaling the prefiller or decoder instance number, many previously explored scaling policies are directly adopted from conventional cloud microservices~\cite{knative-autoscaling,k8s} and monolithic, non-PD disaggregation
LLM serving designs, which also often rely on legacy metrics. 
For example, scaling policies used by AIBrix~\cite{aibrix}, DistServe~\cite{zhong:distserve}, and BlitzScale~\cite{zhang:blitzscale} rely on coarse-grained request counts (e.g., RPS, concurrency) that obscure the fine-grained, token-level bottlenecks that are root causes of performance issues, leading to up to $30\%$ SLO violations and $14\%$ higher GPU costs. Others~\cite{li:flowkv,xu:slinfer-hpca26,yu:faascale-mlsys26,zhu:polyserve} use lagging indicators, such as time-averaged GPU utilization; these react too slowly, scaling up only after a burst has already overwhelmed the system and caused performance degradation. Also, some works~\cite{yousefijamarani2025hyperflexis,du:ecoserve-osdi26,feng:windserve} that use reactive policies trigger scaling only after SLOs have already been breached. These approaches treat the system as a black box, thereby fundamentally misunderstanding its state and leading to poor performance and costly responses.
% \dmi{I don't see a reason why we should go so verbose below. I'd compact it into a single short sentence. Also, I don't think SplitWise's specific mechanism requires an entire sentence here, it seems to break the story} \cz{I classify the SplitWise, ServerlessLLM, and blitscale as the mechanism optimization methods to adapt to the request burst, which is in the same layer of the scaling policy.}

Additionally, to serve sudden request bursts under SLO, prior studies have also explored mechanism optimizations such as rapid instance initialization through accelerated model loading~\cite{fu:serverlessllm,zhang:blitzscale} and instance conversion~\cite{patel:splitwise}. However, the initialization overhead is not limited to model loading, but also includes the non-trivial cost of framework initialization. In contrast, instance conversion~\cite{patel:splitwise} avoids repeated initialization, but naively switching an instance between prefiller and decoder roles can still lead to mutual interference, as the conversion degrades the performance of the instance in its original serving category.

We show that an effective scaling system for PD disaggregation must integrate three co-design components: the leading metric, the stage-aware policy, and the burst adaption mechanism. First, it requires an LLM-native scaling metric that can expose dominant resource bottlenecks across the prefill and decode stages before performance degrades. This metric should also react instantly to traffic changes, providing sufficient time for scaling decisions. Second, the scaling policy must be stage-aware: scale prefillers rapidly to meet TTFT SLOs while precisely managing decoders to sustain TPOT SLOs. Finally, the burst adaptation mechanism must be purpose-built to mitigate the SLO violation caused by the high initialization latency of new LLM instances. 
We collate the state-of-the-art methods in Table~\ref{tab:system_comparison}, listing their lack of the leading metric for scaling and other limitations. 
\begin{table}[t]
    \centering
    \small
    % \resizebox{\linewidth}{!}
    {\begin{tabular}{c|c|c|c}
        \hline
        System& Leading metric & PD stage aware & Rapid scaling  \\
        \hline
        AIBrix\cite{aibrix} &\color{red}\ding{55}&\color{red}\ding{55}& \color{red}\ding{55}\\
        \hline
        DistServe\cite{zhong:distserve} &\color{red}\ding{55}&\color{green}\ding{51}&\color{red}\ding{55} \\
        \hline
        SplitWise\cite{patel:splitwise}&\color{red}\ding{55}&\color{green}\ding{51}&\color{green}\ding{51} \\
        \hline
        BlitzScale\cite{zhang:blitzscale}&\color{red}\ding{55} &\color{red}\ding{55}&\color{green}\ding{51} \\
        \hline
        \textbf{\sys (ours)} & \color{green}\ding{51}&\color{green}\ding{51}&\color{green}\ding{51} \\
        \hline
    \end{tabular}
    % \vspace{-10pt}
    }
    \caption{\sys vs. state-of-the-art systems for scaling PD-disaggregated LLM serving.
    \vspace{-30pt}
    }
    \label{tab:system_comparison}
    
\end{table}
We introduce \sys, an autoscaling framework that resolves the performance mismatch in the PD disaggregation architecture with two innovations, including the Token Velocity and the Convertible Decoder.
To replace lagging and coarse-grained signals, we introduce the \emph{Token Velocity} metric as the foundation of the scaling policy. Token Velocity quantifies an instance’s maximum token-processing capability under its current resource allocation, revealing bottlenecks across different stages. 
Unlike request counts or utilization metrics, Token Velocity is a fine-grained, predictive indicator of system capacity. We define distinct velocity metrics for each stage: prefill velocity, measuring GPU compute throughput; network velocity, capturing KV-Cache~(KVC) transfer rates; and decode velocity, reflecting how quickly decoders release GPU memory.
By monitoring the ratio between incoming token rates and stage-specific velocities, the policy can accurately and proactively scale, preventing SLO violations.

% \dmi{add online profiler}
% \cz{what are these decisions? The Convertible Decoder is only used to absorb the burst, rather than for fast scaling. Make the function of Convertible Decoder clear. }
Additionally, we design the \emph{Convertible Decoder} as the burst adaptation mechanism. This mechanism enables some of the decoders to temporarily operate as prefillers, absorbing traffic bursts by exploiting the observation that both instance types share model weights and that memory-intensive decoders often have spare compute cycles. When a burst arrives, the system immediately routes the excessive prefill requests to the Convertible Decoders.
% \dmi{little overhead is unclear, let's say: immediately instead (i don't see any ovhd there)}
% \sout{excessive prefill requests are routed to the Convertible Decoders \cz{with little overhead, which can avoid the TTFT violation}. }
To prevent resource contention with ongoing decoding tasks, we use an SLO-aware restricted prefill strategy. This involves carefully sizing prefill chunks and reserving a dedicated portion of GPU memory and compute, ensuring that neither the co-located decoding tasks nor the new prefill tasks can violate their TTFT or TPOT SLOs.  After processing the prefill chunks, the instance seamlessly transitions to decoding them. To precisely estimate the appropriate number of Convertible Decoders and avoid interference between prefill and decode, we adapt an online profiler to determine the number of Convertible Decoders based on past traffic. 

We evaluate \sys as a control plane orchestrating a cluster of inference engines deployed on top of vLLM~\cite{kwon:efficient} and LMCache~\cite{liu:lmcache}, when replaying sampled production traces~\cite{choukse:dynamollm} against Llama and Qwen models of various sizes and tensor parallelism degrees on Nvidia A100 and AMD MI308X GPUs. Compared to state-of-the-art baselines, including AIBrix~\cite{aibrix}, BlitzScale~\cite{zhang:blitzscale},  DistServe~\cite{zhong:distserve}, and SplitWise~\cite {patel:splitwise}, \sys outperforms all baselines across both hardware platforms. 
Compared to the above baselines, \sys improves the performance/cost by $1.3$-$6.4\times$, decreasing SLO violation rate from $12$-$87\%$ to $4$-$23\%$ while reducing GPU costs by $0.2$-$13\%$.

% \rui{In summary, this paper makes the following contributions:}
% \begin{itemize}[leftmargin=0cm,itemindent=.4cm,labelwidth=\itemindent,labelsep=0cm,align=left]
%     \item \rui{We characterize the distinct scaling requirements of the prefill and decode stages under bursty traffic, showing that prefillers require rapid scaling while decoders require a leading indicator (\S\ref{sec:back_pd_reqs}).}
%     \item \rui{We propose Token Velocity, a stage-specific capacity metric that translates the arrival-time token demand into prefiller and decoder instance counts, where the decoder demand accounts for the request input/output-length mix (\S\ref{sec:token_velocity}, \S\ref{sec:design_velocity}).}
%     \item \rui{We design the Convertible Decoder, which lets selected decoders temporarily absorb prefill work when prefill demand rises faster than prefillers can be provisioned, within profiled TPOT-interference and memory budgets (\S\ref{sec:convertible}).}
%     \item \rui{We implement \sys atop vLLM and LMCache and show on AMD and Nvidia clusters with production traces that it outperforms AIBrix, BlitzScale, DistServe, and SplitWise in both SLO attainment and GPU cost (\S\ref{sec:e2e}).}
% \end{itemize}

\section{Background and Motivation}

\subsection{LLM Serving Workload}
\label{sec:back_llm_101}

% TODO: Explain the LLM inference process (Prefill vs. Decode), the key metrics TTFT, TPOT, throughput...
LLM inference generates tokens sequentially from input prompts by processing each token through decoder layers with self-attention and feed-forward components. 
% The process ends at a special end-of-sequence token. 
% For efficiency, KV-Cache~(KVC) are stored in GPU memory to avoid re-computing previous tokens.
The LLM inference process can be divided into two phases: the \textit{Prefill} phase and the \textit{Decode} phase. During the prefill phase, KVCs are constructed for all input tokens, which is a highly-parallel compute-intensive process. The subsequent decode phase is autoregressive, sequentially generating output tokens.
% incurs substantial computational overhead. In the subsequent decode phase, the model generates output tokens autoregressively. 
In each iteration, it reuses the KVC of previously generated tokens and computes only the KVC for the last token before sampling the next output token. This process can result in substantial memory consumption, but it is much less compute-intensive than the prefill phase.
% incurring only minimal computational overhead.
LLM serving systems can be characterized with two key SLOs:
% \sout{end-to-end response time}\cz{We don't measure e2e latency, remove it}, 
TTFT and TPOT characterized by the execution of the prefill and decode phases, which evaluate the system's interactivity and the fluency of output generation~\cite{choukse:dynamollm,zhong:distserve}.

% Since the inference process comprises two phases, LLM serving performance is typically characterized not only by traditional metrics such as throughput and end-to-end latency, but also by the TTFT in the prefill phase and the TPOT in the decode phase.

\begin{figure}[tbp]
  \centering
  \includegraphics[width=\linewidth]{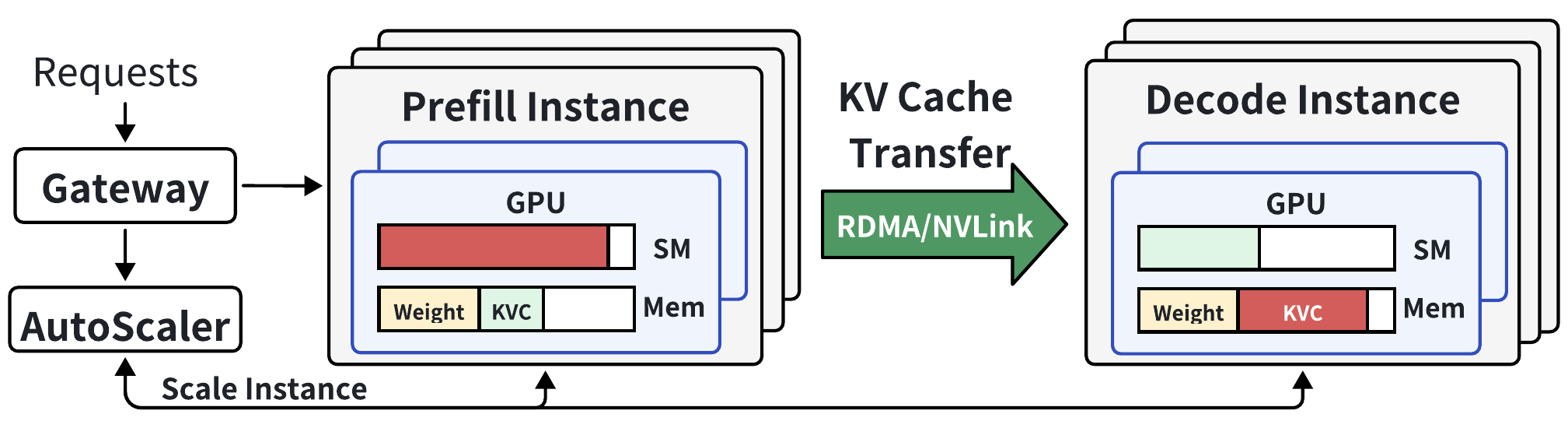}
  \caption{Disaggregated LLM serving system architecture.
  \vspace{-8pt}
  }
  
\label{fig:pd_architecture}
\end{figure}
\subsection{Serving LLM with PD Disaggregation}
\label{sec:back_pd_101}

Due to the differences between the prefill and decode phases, particularly in their compute and memory requirements, batching requests across the two phases may cause performance interference~\cite{zhong:distserve,patel:splitwise}. Therefore, many recent systems~\cite{zhong:distserve, patel:splitwise, qin:mooncake} disaggregate the prefill and decode phases, executing them on different GPUs. As shown in Fig.~\ref{fig:pd_architecture},
when a request arrives, it is first sent to a prefiller. After the prefiller computes the KVC of the prompt, it sends the KVC to a decoder via high-speed interconnects such as RDMA or NVLink.  
% \dmi{\sout{Once the transfer is complete, the decoder, begins autoregressively generating new tokens using the received KV cache.} 
% Using the received KVC, the decoder instance generates output tokens.} 
Using the received KVC, the decoder generates output tokens.
% In this architecture, the compute-intensive prefill phase can be separated from the memory-intensive decode phase. As a result, 
By separating the compute-intensive prefill phase from decode, both TTFT and TPOT become more stable.
% \sout{, making the system more interactive and fluent.}
% the performance of the decode phase remains stable without stragglers in GPU computation resources. At the same time, disaggregating the prefiller and decoder makes it easier to adjust the number of instances based on their own resource requirements, providing greater scaling flexibility.

Furthermore, the PD disaggregated architecture enables
% \sout{Another benefit of the PD disaggregation is in enabling} 
flexible pooling and autoscaling of GPU resources allocated for prefiller and decoder instances. While prior works~\cite{zhong:distserve,patel:splitwise,qin:mooncake} have showcased the performance advantages of the PD disaggregation, defining the workload requirements and system implications for PD system autoscaling is key to achieving cost-efficient operation under the strict SLOs. 

% \dmi{Need to mention Mooncake and SplitWise. Particularly, SplitWise's re-purposing decoders for prefillers}

\begin{figure}[t]
    \centering
        \includegraphics[width=1\linewidth]{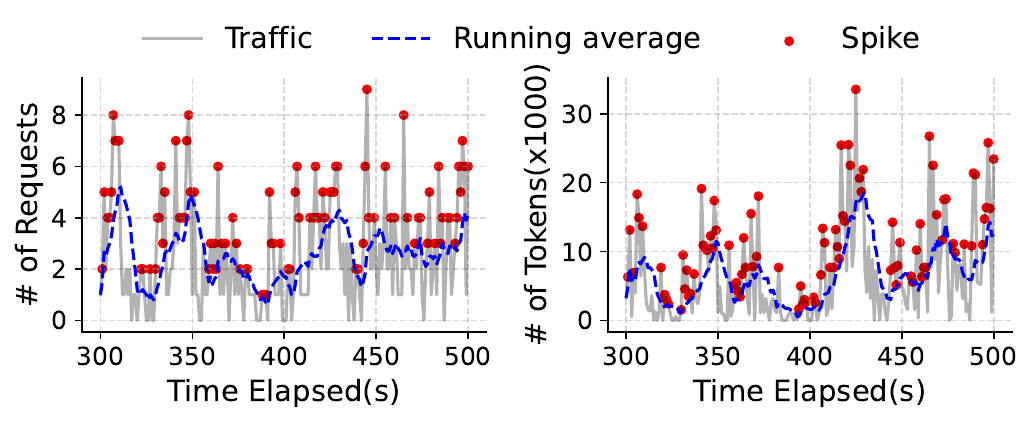}
    \caption{
    Traffic represented as requests (left) and tokens (right) in a production code trace~\cite{choukse:dynamollm}. Bursts are the spikes above the running average.
    \vspace{-10pt}
    }
    \label{fig:burstiness}
\end{figure}

\subsection{Workload Implications for PD Disaggregated Systems}
\label{sec:back_burst_wld}

\label{sec:motiv_burst}
\begin{figure}[t]
    \begin{subfigure}[b]{0.49\linewidth}
        \centering
        \includegraphics[width=1\linewidth]{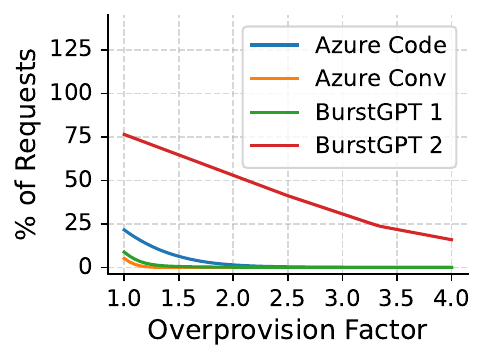}
        \caption{Request bursts percentage.\vspace{-10pt}}
        \label{fig:over_provision}
    \end{subfigure}
    \hfill % Adds horizontal space between subfigures
    \begin{subfigure}[b]{0.49\linewidth}
        \centering
         \includegraphics[width=1\linewidth]{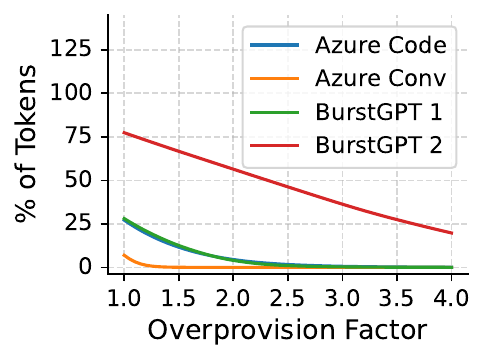}
        \caption{Token bursts percentage.\vspace{-10pt}}
        \label{fig:burst_token}
    \end{subfigure}
    \caption{Percentage of burst traffic while varying the overprovisioning ratios in four production traces~\cite{choukse:dynamollm,wang2025burstgpt}. 
    \vspace{-15pt}
    }
    \label{fig:burst_workload}
\end{figure}

% \hong{better to unify the name for PD achitecture/disaggregation}

% \dmi{replace the parameter with discussing the overprovisioning directly}
\subsubsection{Bursty traffic in production.}
% LLM serving workloads from real-world applications exhibit extremely bursty behavior~\cite{choukse:dynamollm, xu:llmmesh}. 
Similar to other cloud applications, LLM serving workloads often exhibit highly bursty behavior, as shown by prior studies of production traces~\cite{choukse:dynamollm,xu:slinfer-hpca26}.
% \rui{The following section is completely re-written}
In LLM inference workloads, bursts can occur along two dimensions: requests per second (RPS) and input tokens per second (TPS).
% \hong{better to clarify input token? throughput is also tok/s}.
We examine both types of bursts using production inference traces from Azure~\cite{choukse:dynamollm} and BurstGPT~\cite{wang2025burstgpt}.
% and analyze why naive overprovisioning fails to handle them effectively. 
We apply a 1-minute sliding window, same as \cite{choukse:dynamollm}, to compute the average request and token rates, assuming a system that gradually scales following the running average can deliver a throughput equal to these long-term average rates. 
% \dmi{unclear why 1 min, refer to the scaling time of the existing works, e.g., dynamollm}
Fig.~\ref{fig:burstiness} illustrates that, when the system serves Azure conversational trace, a significant fraction of traffic comes as \emph{bursts}, exceeding the running-average trendline, causing queuing and resource contention.

% such a system has to serve a large fraction of traffic as bursts that exceed its current serving capability.\footnote{We define a \emph{burst} as the traffic fraction of requests, or tokens, that exceeds the running average trendline.}}
% \cz{This para only discusses the efficiency of the overprovision strategy, without cost.}
Next, we evaluate whether static overprovisioning of cluster resources can avoid the contention caused by the burstiness.
We model an autoscaling system that allocates $X$ times more resources, hence delivering $X$ times more throughput than the running average, sweeping $X$ from 1 to 4. Fig.~\ref{fig:burst_workload} shows the percentage of requests or tokens exceeding this provisioned throughput under different overprovisioning factors. Clearly, naive resource overprovisioning cannot handle bursty traffic. For example, Fig.~\ref{fig:over_provision} shows that the BurstGPT~2 trace has around $25\%$ of requests beyond the throughput of a $3\times$ overprovisioned system. Fig.~\ref{fig:burst_token} reveals the same conclusion in terms of token arrival rates.
Thus, while helpful, overprovisioning alone is not a panacea for LLM serving clusters.

\begin{figure}
    \centering
    \begin{subfigure}[b]{0.49\linewidth}
        \includegraphics[width=1\linewidth]{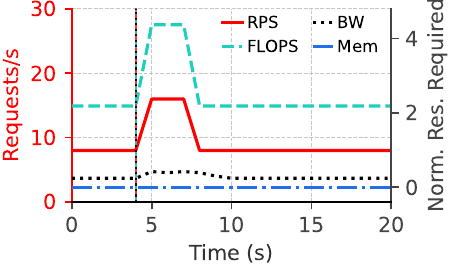}
    \caption{Prefiller.}
    \label{fig:prefiller_behavior}
    \end{subfigure}
    \begin{subfigure}[b]{0.49\linewidth}
        \includegraphics[width=1\linewidth]{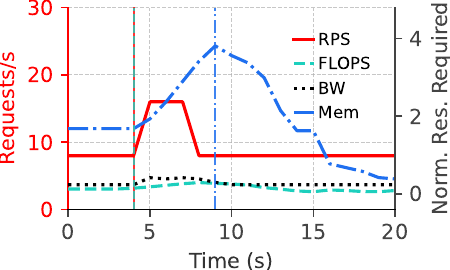}
    \caption{Decoder.}
    \label{fig:decoder_behavior}
    \end{subfigure}
    \caption{
    Utilization of compute~(Comp) and memory capacity~(Mem) on a single-GPU Prefiller and Decoder instances, and network bandwidth~(Net) normalized to their maximum values in the A100 cluster, when serving a burst of requests with \llama. 
    \vspace{-15pt}
    % \TODO{Comm to Net}
    % The BW is normalized by the bandwidth of an RDMA. 
    % \dmi{The right axis' label to "Norm. utilization" from 0 to 1}
    }
    % \dmi{RDMA bw is too much detail for this caption}
    \label{fig:pd_behavior}
\end{figure}

% \subsubsection{Scaling Requirement of PD Disaggregation in Burstiness}
\subsubsection{Scaling requirements for PD stages in the presence of bursts.}
\label{sec:back_pd_reqs}

% \dmi{R3 leading vs lagging}
% \rui{TODO: need to explitcitly points out that prefiller locates at upstream of the pipeline, while decoder locates at downstream, therefore has different scaling requirements.}

To avoid overprovisioning, we need to define the requirements for cost-efficient dynamic resource scaling for both prefiller and decoder instances, given their resource usage differences (\S\ref{sec:back_llm_101}). We conduct an experiment showing the time between burst arrival and SLO violations, which we refer to as \textit{reaction window}.
We set up two prefillers and one decoder in the A100 cluster~(setup details are in \S\ref{sec:method}), hosting a \llama model. We simulate a bursty workload that starts with stable RPS=8 and changes to RPS=16 at 4s, which lasts for 4 seconds before returning to the stable traffic. 
% The experiment is conducted on A100-40GB GPUs, interconnected by 200Gbps RDMA. \TODO{just refer to the A100 cluster and leave the details to Eval}
% We collect compute (FLOPs), network(BW), and memory (Mem)  \dmi{no need for FLOPS, BW, etc. in the text, this is in the caption} resource utilization on prefillers and decoders. Based on the results, we demonstrate different resource usage characteristics for the prefiller and decoder.
% \TODO{using the following methodology (a couple of sentences about the microbenchmark and metrics in Fig.~\ref{fig:decoder_behavior}.}

\noindent\textbf{Prefiller stage.}
As shown in Fig.~\ref{fig:pd_behavior}, when PD disaggregated LLM serves the bursty workload, the prefiller encounters the burst first. The prefiller requires more GPU computation resources to handle the request immediately, while network bandwidth and GPU memory demands remain stable.
Such an immediate change in compute demand results in the TTFT violations almost immediately, leaving a limited reaction window for the prefiller to handle the burst. Such a contradiction necessitates the need for a specifically designed burst adaptation mechanism.
% \cz{We don't design any cold start mechanism, better not to mention it.}
% During the prefill phase, requests typically exhibit stable network bandwidth and GPU memory demands. This stability arises from the fact that the prefiller operates with a small batch size (often set to 1~\cite{du:prefill_only}), as the prefill stage is compute-intensive ~\cite{choukse:dynamollm}.
% Therefore, since a spike in compute usage arises immediately after a burst arrival, the first scaling requirement for the PD disaggregation system is:

% \dmi{tbh, the above text doesn't really lead to the below conclusion as of now. it does start with the right statement but then the flow diverges into stability of net and mem and I found myself lost when I got to R1. need a bridge sentence}

\reqbox{%
% \textbf{{R1: The rapid increase in resource demand necessitates rapid scaling of the prefiller to satisfy the SLO for TTFT.}}
\textbf{{R1: Located upstream in the PD pipeline, the prefiller demands rapid scaling to handle sudden resource bursts}}
}

\noindent\textbf{Decoder stage.} As illustrated in Fig.~\ref{fig:decoder_behavior}, the decoder’s resource demand increases with a distinct delay, contrasting with the prefiller's immediate surge. The demand sequence begins with network bandwidth, followed by compute resources and GPU memory. Notably, the decoder's memory consumption does not peak immediately following the burst. This lag allows the decoder to react in time, provided the autoscaler can foresee the incoming traffic surge. Such behavior emphasizes the necessity of an accurate leading indicator to reflect shifts in incoming traffic.

\reqbox{%
% \textbf{R2: Scaling the decoder can tolerate delays of up to a few seconds, but it requires accurate decisions made beforehand to avoid performance degradation.}
\textbf{R2: Decoder locates at the downstream of the PD pipeline. It requires a leading indicator that forecasts future trends to react accurately.}
% \colin{What is second-level delay? Do you mean delays in seconds?}
}
% \cz{discuss the resource requirement of the prefiller and decoder under the burstiness. Wish to get two main takeaways: 1. Prefiller requires scale immediately to avoid the SLO violation on TTFT, 2. Decoder can be scaled with a delay, and it can be determined whether it should be scaled when the request burst arrives due to the linear relationship between the request number and the memory requirement.}

% In addition to their \emph{high frequency}, real-world workload bursts also exhibit \emph{short durations}. As shown in Fig.~\ref{fig:spike_duration}, the average burst duration is less than 2 seconds. Furthermore, for most traces, the average burst duration decreases rapidly as the burst threshold increases. This indicates that setting a higher threshold results in capturing more rapidly fluctuating burst events, which are shorter and harder to handle effectively.

\subsection{Limitations of Existing Work}
% \colin{Limitations* (plural)?}
\label{sec:back_sota_limits}

% TODO: Provide quantitative motivation using workload traces.
\subsubsection{Inaccurate and delayed scaling policy.}
Over the last decade, researchers have explored designs of serverless systems, such as AWS Lambda~\cite{aws-lambda-scaling}, Google Cloud Run~\cite{googlecloudrun}, Knative~\cite{knative-autoscaling}, which scale underlying resources following traffic changes in microservices and serverless applications.
% \dmi{There must be a lead-in sentence: Over the last decade, researchers have explored designs of elastic systems, such as AWS Lambda, Google Cloud Run, Knative, \dmi{cite them} which scale underlying resources following traffic changes in microservices and serverless applications.}
Many existing LLM systems~\cite{aibrix,zhong:distserve,xiang:aegaeon, zhang:blitzscale} directly inherit the scaling approaches from these serverless systems, retrofitting their policies and mechanisms, which we classify in the following three categories: 
% inherit part of the designs in conventional CPU-tailored scaling systems, overlooking characteristics in PD disaggregated systems. 
% We classify these approaches into three categories: Request-based, utilization-based, and performance-based.
% However, scaling policies used in conventional CPU-tailored systems are ill-suited for PD disaggregated LLM inference workloads. 
% \dmi{the below phrasing is odd: current policies cannot address ... because? is it bc they don't account for GPUs?}
% are hard to address this burstiness in current PD disaggregation systems.
% ~\cite{zhong:distserve,aibrix, zhang:blitzscale}:
% \begin{itemize}[leftmargin=*]
    % \item 
% \rui{I suggest divide in three types: First. Concurrency based policy: AIBrix's perfiller, Blitzscale's prefiller and decoder. Second, RPS based policy: Distserve. Its problem is: rps cannot accurately reflects resource bottleneck, traffic pattern would affect the scaling thresholds. Third, Resource utilization based scaling policy. AIBrix's decoder}
% \cz{Request-based includes the waiting requests in the queue(Concurrency) and the number of the requests (RPS)}

% \dmi{R, please proofread the rest of this sec. the writing is good but verify that it covers all the relevant works}\TODO{}\\

\noindent\textbf{Request-based scaling policies} are widely used by current PD disaggregation systems, such as AIBrix~\cite{aibrix}, DistServe~\cite{zhong:distserve}, BlitzScale~\cite{zhang:blitzscale}, and Aegaeon~\cite{xiang:aegaeon}. Based on the point of measurement, request-based policies are categorized into two types: concurrency-based policies, which track the number of requests actively in the queue, and RPS-based policies, which track the rate of requests entering the system. The concurrency-based policy triggers scaling by checking whether the number of requests in the queue within a sliding window exceeds a specified threshold~\cite{aibrix,zhang:blitzscale}. However, when the system experiences a burst of requests, changes in concurrency occur with a delay, which violates the requirement for scaling prefillers \textbf{R1}. This is because the sliding window averages out bursty traffic through overlapping requests, thereby reducing its apparent impact on overall utilization. RPS-based policies~\cite{zhong:distserve} trigger scaling when the RPS exceeds a predefined threshold. For example, DistServe derives such thresholds from offline-profiled static throughput estimates, which do not capture the runtime demand variation of decoders across workloads with different input/output-length distributions. However, RPS alone cannot accurately capture resource bottlenecks, leading to suboptimal decisions in determining the number of decoder instances and thus violating the scaling requirement \textbf{R2}.

\noindent\textbf{Utilization-based scaling policies} monitor the resource utilization of the PD disaggregation system, including both the resource utilization of the instance in the PD disaggregation system and LLM serving performance~\cite{aibrix,li:flowkv,xu:slinfer-hpca26,yu:faascale-mlsys26,zhu:polyserve}. For example, AIBrix~\cite{aibrix}, in its HPA, KPA, APA policies adopted from Knative~\cite{knative-autoscaling}, scales instances by monitoring the average GPU memory utilization across all instances within the window and triggers the scaling when the utilization exceeds the threshold. However, the lag in GPU resource utilization, particularly in memory usage, prevents utilization-based scaling policies from accurately determining both the necessity of scaling and the appropriate number of instances. Hence, such policies risk violating the stringent prefiller requirements \textbf{R1} while simultaneously failing to accommodate the demands of the decoder \textbf{R2}.

\noindent\textbf{Performance-based scaling policies} target overall system performance within PD disaggregation systems, considering metrics such as TTFT and TPOT~\cite{yousefijamarani2025hyperflexis,du:ecoserve-osdi26,feng:windserve}. However, these systems trigger scaling only after SLOs are violated, which is much later than resource shortage manifests. 

\subsubsection{Inefficient burst adaptation mechanisms.}
\label{sec:split_wise}
 
Current work also tries to adapt to the burst workload with two main mechanisms of optimization: 1) quick initialization of the instance~\cite{zhang:blitzscale,fu:serverlessllm} and 2) the instance role conversion~\cite{patel:splitwise}.
% \dmi{Above is a complex and unclear sentence, it can be phrased better}
For the former, \sllm~\cite{fu:serverlessllm} proposes caching model weights in the server's host memory, accelerating model loading speed; \blitz~\cite{zhang:blitzscale} proposes parallelizing model loading with computation, allowing instances to start processing requests; before the model weights are fully loaded.
Medusa~\cite{zeng:medusa} accelerates startup by restoring offline-materialized KV-cache state and CUDA graphs to avoid expensive runtime profiling and graph capture.
These methods, nevertheless, leave the remaining initialization delays on the critical path.
% , rendering them inefficient to meet the tight TTFT and TPOT SLOs for large models.
% \cz{However, 
% \sout{the initialization process of the instance includes model weight loading, but also the engine initialization and CUDA graph capturing.}
% focusing solely on model weight loading overlooks the rest of the initialization pipeline, which can take several seconds and subsequently severely increases TTFT during traffic bursts and violates the scaling requirement \textbf{R2}.}
% Depending on the model size and tensor parallelism, this process can take many seconds, severely degrading the TTFT during traffic bursts. Recent works have explored various strategies to mitigate this overhead.
% \dmi{unclear which overhead}
%  However, these two approaches focus strictly on model loading and ignore the rest of the initialization process, such as engine initialization and  CUDA graph capturing. 
% \dmi{which ones?}
% \rui{Rewrite}
For the latter, \spw~\cite{patel:splitwise} proposes a mechanism to convert existing prefill instances into their counterparts, and vice versa, to absorb sudden requests and token bursts. However, this method may lead to substantial interference between co-resident prefill and decode tasks,
% and hence violated TTFT and TPOT SLOs, 
as we reveal in \S\ref{sec:e2e}, because the method relies on conversion as a best-effort fallback rather than a precise scaling mechanism. Moreover, SplitWise neither bounds the interference introduced by conversion nor distinguishes the different burst characteristics of the prefill and decode stages.
Thus, none of these works manages to fulfill \textbf{R1} or \textbf{R2}.

\section{Core Methods}
\label{sec:core_method}
In this section, we introduce the Token Velocity metric and the concept of Convertible Decoders to address each stage's scaling requirements in a PD disaggregation system.

\subsection{Design Principles}
% The design of core methods is guided by two key principles, derived from the scaling requirements of the prefiller and decoder:
We start by formulating two design principles based on the scaling requirements discussed in \S\ref{sec:back_pd_reqs}.

\subsubsection{Accurately balancing the PD resources.}
% in advance
This principle is applied to guide the scaling of the decoder (\textbf{R2}). 
% \hong{why first introduce R2 not R1?} 
% \dmi{The number of requests received by prefillers subsequently causes  an increase in the resource demands for decoders down the PD pipeline.
% At the same time, if decoders' resources saturate, decoders backpressure the prefillers, which become unable to send KV cache data at the same rate. Finally, the same applies to the network bandwidth connecting prefillers and decoders as a middle stage of the PD pipeline.
% }
The number of requests received by prefillers subsequently causes an increase in the resource demands for decoders down the PD disaggregation system.
At the same time, if decoders' resources saturate, decoders backpressure the prefillers, which become unable to send KV-Cache~(KVC) data at the same rate. Finally, the same applies to the network bandwidth connecting prefillers and decoders as a middle stage of the PD disaggregation system.
% imposes the growth of decoder resource demand.
% When the decoder’s available resources fall short of the requirements imposed by these requests, its performance becomes the bottleneck of the entire system. Conversely, if decoder resources exceed the actual demand, unnecessary overhead is introduced. 
Hence, the system must proactively balance decoder resource allocation with the number of active prefillers to avoid both performance degradation and resource waste.

\subsubsection{Timely scaling tailoring to prefiller rapid resource demand bursts.} 
% This principle is applied to address the scaling requirements of the prefiller (\textbf{R1}). 
Scaling LLM inference engines often violates the \textbf{R1} requirement for rapid prefillers scaling, incurring high start-up overheads to allocate GPU memory, load model parameters and initialize the runtime.
%, and initializing the model.
    % \cz{To Ruiqi: find a reference to support this sentence}
This process typically requires $3$-$10$~seconds, depending on the model size and tensor parallelism level~\cite{zeng:medusa}, significantly degrading prefillers' performance, i.e., TTFT, in the presence of bursty request and token rates. 
    % Such delays can significantly degrade prefiller performance under bursty workloads. 
Thus, the system must minimize the prefiller start-up delays to meet the TTFT SLO.
% ensure that the prefiller attains its SLO.
    % This challenge stems from the fact that different requests introduce various resource requirements and finally impact the overall system capacity. Typically, the maximum request number that can be handled by the system (system capacity) is constrained by the most limited resource. However, because multiple resources are involved in the inference process, the resource demands of requests are highly dynamic. For example, the unpredictable output length makes the decoder's memory usage variable, while load balancing strategies introduce fluctuations in network bandwidth and the computing demand on the prefiller. As a result, dynamically identifying system resource bottlenecks becomes difficult, which in turn complicates the quantification of system capacity.

\begin{figure}
    \centering
    \includegraphics[width=1\linewidth]{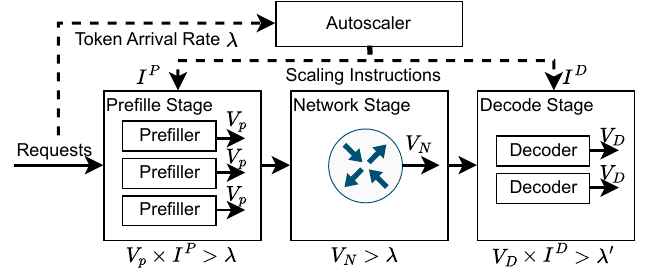}
    \caption{Overview of Token Velocity in prefill, network, and decode stages. $\lambda^{\prime}$ denotes the combined rate of input and predicted output tokens. Token velocity denotes the maximum token processing rate at the prefill~($V_P$), network~($V_N$), and decode~($V_D$) stages. The autoscaler adjusts the number of prefillers~($I^P$) and decoders~($I^D$) based on the ratio between the token arrival rate and the corresponding stage velocity, ensuring that no stage becomes a bottleneck.}
    % \TODO{Replace $I^X_r$ with $I^X$ in the fig}
    % }
    \vspace{-10pt}
    \label{fig:token_velocity}
\end{figure}

% \begin{figure}
%     \centering
%     \includegraphics[width=1\linewidth]{figures/design/tokenvelocity.pdf}
%     \caption{The example of the calculation of output velocity, with the maximum output velocity highlighted in red. The system can process 3 and 4 tokens per second during the prefilling and network stages, respectively, while the GPU memory can accommodate 7 tokens during the decoding stage.}
%     \label{fig:token_velocity}
% \end{figure}
\subsection{Token Velocity}
\label{sec:token_velocity}

% \dmi{Add: we introduce a \textbf{leading metric}, i.e., Token Velocity}

% \rui{Need to mention we introduce a leading metric, i.e. Token Velocity}
% To determine whether scaling is necessary, the most straightforward approach is to examine whether the current system has sufficient resources to accommodate incoming requests within the SLO requirement. If not, the system must be scaled out. 
% However, a traditional resource utilization-based metric is difficult to reflect the system's capacity for two main reasons. First, resource utilization often exhibits a significant delay. For instance, when a new request requires prefilling, continuous batching forces the request to wait until the ongoing iteration completes before prefilling can begin. As a result, if prefillers rely on the current GPU compute utilization as their baseline metric to scale, the prefiller’s reported GPU utilization does not accurately reflect the actual availability at the start of the new request, which can easily lead to overestimation or underestimation of the available resources.
% Second, resource utilization ...

% \subsubsection{Output Velocity}

% \dmi{this is the only place we mention Output Velocity (e.g., Fig 5 doesn't show it). I think it should be enough to directly define Token Velocity as maximum t/s for a stage.}

To apply the first principle, 
% \sout{we must accurately balance PD resource requirements in advance. \rui{Add a sentence.} To achieve this}, 
we introduce a leading indicator \emph{Token Velocity}: the maximum number of tokens an instance can output per second. Traditionally, evaluating system bottlenecks is difficult because prefillers, networks, and decoders operate at different speeds. Token Velocity solves this by acting as a universal common denominator that unifies prefill ingestion, network transfer, and decoder output speeds. Consequently, by measuring the incoming token arrival rate against this unified Token Velocity, the system can instantly calculate the precise number of instances needed, allowing it to react to traffic bursts in real time.
% Consequently, by measuring the incoming token rate against Token Velocity, the system can instantly calculate the precise number of instances required to react to traffic bursts timely.

Due to the differences in resource requirements in each stage of the PD disaggregation system~(\S\ref{sec:back_pd_reqs}), we calculate Token Velocity for each stage with a different method.
% Due to the differences in resource requirements for different phases in the PD disaggregation framework~(\S\ref{sec:back_pd_reqs}), the approach to \dmi{estimate} Token Velocity of different stages is different.
% To accurately balance the PD resource in advance, we introduce a new metric called \emph{Token Velocity}, which represents \textit{the maximum number of tokens that can be released from the instance in a second with the current allocated resource}, for identifying the resource bottleneck of the system.
% When requests A and B arrive one by one, the prefiller starts to prefill each request according to the arrival order. Due to the limitation of the GPU computation capacity, the prefiller should handle only three tokens at a time. Therefore, the request A can be handled in only one second, and after that, the prefilled request is asynchournously sent to the network transmission phase (Network Stage) to deliver to the decoder. Then the prefiller starts to handle the request B which exceeds the maximum number of tokens that the prefiller can handle.

\begin{itemize}[leftmargin=0cm,itemindent=.4cm,labelwidth=\itemindent,labelsep=0cm,align=left]
    \item \emph{Prefill Velocity ($V_P$)} 
    is the maximum speed at which prefillers can process input tokens. $V_P$ is bound by GPU compute and is constant for a combination of a GPU generation and LLM model. 
    % The primary resource requirement by requests on the prefiller is GPU computation.
    % \dmi{Prefillers are bound by GPU compute, hence} Token Velocity of the prefiller is defined in terms of the number of input tokens the GPU can process per second, which remains constant for a specific model and GPU.
    \item \emph{Network Velocity ($V_{N}$)} is the maximum transmission speed at which prefillers can transfer KVC to decoders, typically over RDMA network or NVLink. $V_N$ depends on network bandwidth and KVC size.
    % In a PD disaggregation framework, transmitting KVCache populated during the prefill phase to the decoder entails transferring data over links, such as the network, RDMA, or PCIe. 
    % To characterize this, we define Token Velocity as the maximum token transmission rate that can be sustained without incurring significant congestion on the transmission link with minimum bandwidth.
    \item \emph{Decode Velocity ($V_D$)} 
    is the maximum rate at which decoders drain tokens, measured as the number of tokens completed per unit time when the decoder is saturated, which can be expressed as: 
    \begin{equation}
\small
    V_D=\frac{\sum_{r\in R}L_r}{TPOT\times L_o},
    \end{equation}
    where $R$ is the set of completed requests, $L_r$ is the number of tokens in request $r$, $L_o$ is the average output length,
    and $TPOT$ represents the measured Time-Per-Output-Token metric. It differs from generation throughput: from a memory perspective, generation throughput measures how quickly GPU memory is allocated during token generation, whereas Decode Velocity reflects how quickly memory is released as tokens are finalized. The decoder’s Token Velocity is bounded by available GPU memory capacity. When the memory is full, the decoder will backpressure the prefiller and network stages. This metric connects decoders with the earlier stages, ensuring that input and output rates remain balanced, preventing bottlenecks in the decode stage. Results in \S\ref{sec:utilization} confirm that Decode Velocity is an effective metric for capturing decoder-side pressure.

\end{itemize}

% When the computation volume requirement of the prefilling token is lower than the GPU computation volume, the GPU can handle them immediately rather than incuring a cultimative waiting delay. For example, the request A in Fig.~\ref{fig:token_velocity} can be handled by the GPU within one second in the prefill stage, and the request B can continuously be handled by the GPU as the GPU resource is released. However, since the request B exceeds the maximum number of tokens that the GPU can be handled, it should spend more than one second to complete, leading to the potential risk of the delay of the next request. 

% \subsubsection{Token Velocity}
\label{sec:token_velocity_design}
\begin{figure}[t]
  \centering
  \includegraphics[width=\linewidth]{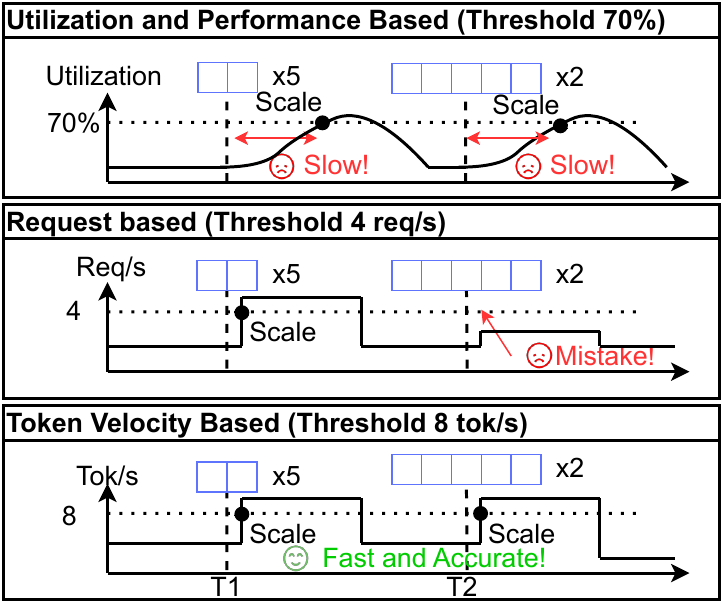}
  \caption{Comparison of scaling policies. Two traffic bursts occur at times $T_1$ and $T_2$: at $T_1$, five requests with two tokens each; at $T_2$, two requests with five tokens each. In both cases, each instance has a Token Velocity of 8 tokens/s, but only the Token Velocity-based policy responds promptly and accurately to both bursts.
  \vspace{-10pt}
  }
  \label{fig:design_policy}
\end{figure}
As illustrated in Fig.~\ref{fig:token_velocity}, the autoscaler monitors the token arrival rate $\lambda$ and the token velocities $(V_P, V_N, V_D)$ at each stage, dynamically adjusting the number of prefiller and decoder instances ($I^P$ and $I^D$ are the required number of prefillers/decoders) to ensure that no stage becomes a bottleneck. For the prefill and network stages, Token Velocity remains constant because their throughput is primarily bounded by GPU computation and network bandwidth. In contrast, the decode stage exhibits variable Token Velocity, as the decoder’s output speed depends on the input and output token lengths of each request. Once these lengths are determined, the decoder’s velocity can be approximated as constant.  To estimate the required number of decoders, incoming requests are classified into different types based on their input and output lengths, and the instance number for each type is computed as the ratio between its incoming token rate and profiled Token Velocity. Summing these per-type requirements yields the number of decoders.  
% \sout{We validate this approach in Sec.~\ref{sec:utilization}. }

% For the decode stage, to accommodate the dynamic variation in output sequence lengths, which causes fluctuations in output velocity, we employ an auxiliary model to classify sequences into different length categories. Requests are then routed accordingly, ensuring that each instance handles no more than two length categories.

% If the aggregate token demand exceeds Token Velocity of a phase, that phase becomes the bottleneck, and scaling is required. 

{Fig.~\ref{fig:design_policy} illustrates why Token Velocity-based scaling policy can outperform existing policies~(\S\ref{sec:back_sota_limits}) with an example of two bursts arriving into a system serving stable inference traffic. In this example, the first is a request burst that carries many requests with a few tokens, while the second is a token burst that carries a few requests with many tokens. 
% For the first request burst (at $T_1$ in Fig.~\ref{fig:design_policy}), 
At $T_1$, a utilization-based policy 
% \TODO{refer to a concrete policy in \S\ref{sec:back_sota_limits}}
reacts slowly to both bursts. As discussed in \S\ref{sec:back_sota_limits}, the resource utilization has a significant lag between the traffic spikes. Such lag would cause a high scaling latency and lead to SLO violations. 
% \dmi{WHY?} 
% A request-based policy responds to the request burst when the number of newly arrived requests exceeds the predefined threshold (e.g., 4 requests per second here). However, for the \sout{request} token burst at $T_2$, the request-based policy makes a mistake because though the number of requests is lower than the threshold, the number of tokens (10 at $T_2$) that should be handled is over the maximum token processing speed of the instance (8 tokens/s), which may lead to the requirement of the instance scaling. 
A request-based policy reacts to bursts only when the arrival rate exceeds a predefined threshold (e.g., 4 req/s here). However, at time $T_2$, the burst is token-intensive rather than request-intensive. Although the number of requests remains below the threshold, the total token volume (10 tokens at $T_2$) exceeds the instance’s maximum processing rate of 8 tokens/s. As a result, the request-based policy underestimates the actual workload and fails to trigger scaling.
In contrast, a Token Velocity-based policy correctly detects both request and token bursts and swiftly and accurately scales the instances. 
% prefillers and decoders.
% that the token arrival rate exceeds its threshold in both bursts and therefore scales swiftly and accurately.

\begin{figure}[t]
  \centering
  \begin{subfigure}[t]{0.49\linewidth}
    \includegraphics[width=\linewidth]{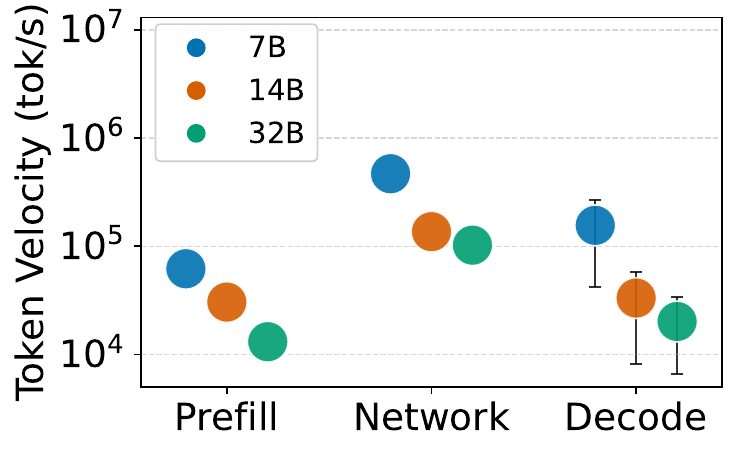}
    \caption{A100 cluster.\vspace{-10pt}}
    \label{fig:velocity_a100}
  \end{subfigure}
  \hfill
  \begin{subfigure}[t]{0.49\linewidth}
    \includegraphics[width=\linewidth]{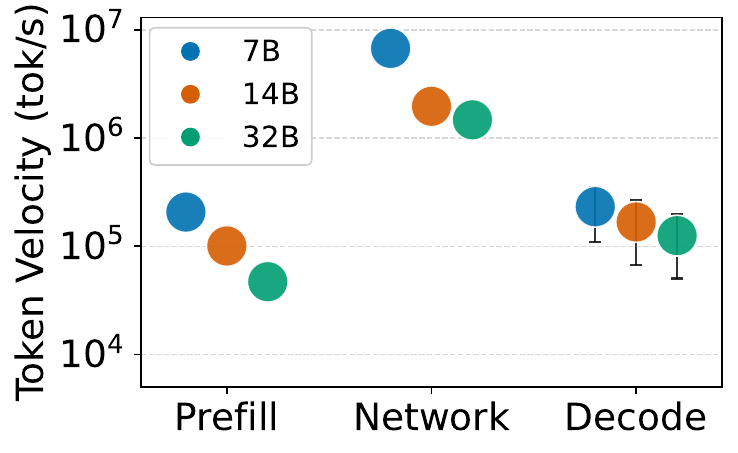}
    \caption{H100 cluster.\vspace{-10pt}}
    \label{fig:velocity_h100}
  \end{subfigure}
  \caption{
  % \sout{Maximum t}
  Token Velocity of prefill, network, and decode stages of Qwen models (7B, 14B, 32B) in A100 and H100 GPU clusters.
  % : A100 cluster and H100 cluster.
  % \sout{the A100 cluster with 4 NVIDIA-SXM-A100-40GB GPUs and 200Gbps RDMA bandwidth; the H100 cluster with 8 NVIDIA-Superpod-H100-80GB GPUs and 2880Gbps RDMA bandwidth.} 
  % \dmi{Call them H100 and A100 clusters, too much detail for a caption}
  Decoder velocity varies with input length and concurrency (indicated with error bars).
  \vspace{-15pt}
  }
  \label{fig:design_velocity}
\end{figure}

% \begin{figure}[t]
%     \centering
%     \includegraphics[width=\linewidth]{figures/evaluation/general.pdf}
%     \caption{Token velocity across different stages and hardware platforms. All models belong to the Qwen-2.5 family. The decoder request types correspond to long input–long output (LL), long input–short output (LS), short input–short output (SS), and short input–long output (SL).}
    
%     \label{fig:general}
% \end{figure}
% \rui{Below paragraph is completely re-written.}
\subsection{Token Velocity Characterization in LLM Deployments}
\label{sec:core_velocity_study}
% \rui{Re-written}
Token Velocity serves as a unified metric for characterizing resource bottlenecks across the prefill, network, and decode stages, enabling system architects to quantify and compare token throughput under different models and hardware configurations. 
In this section, we systematically characterize Token Velocity across these three stages under various model sizes and GPU clusters. 

As shown in Fig.~\ref{fig:design_velocity}, we evaluate three Qwen-2.5 models: (7B, 14B, and 32B) on two GPU clusters: an A100 cluster with 4 A100-40GB GPUs and 200~Gbps RDMA; an H100 cluster with 8 H100-80GB GPUs and 2880~Gbps RDMA.  
For each configuration, we plot Token Velocity of the prefill, network, and decode stages, assuming all GPUs within a node are fully utilized for a single stage. 
For the decode stage, whose Token Velocity depends on request and token patterns, we use error bars to represent its range of variation. 
Across all settings, the network bandwidth is significantly higher than Token Velocity of both prefill and decode stages, suggesting that while prefillers and decoders require careful scaling to maintain balanced performance, network bandwidth does not become the provisioning bottleneck in these setups. In production deployments where bandwidth is constrained or shared, the network can become a bottleneck; \sys accounts for this case by bounding prefiller capacity with the network velocity (\S\ref{sec:design_prefiller_as}).

\subsection{Convertible Decoder}
\label{sec:convertible}

% As the above study revealed, prefillers have the highest likelihood of becoming a bottleneck, given their upstream location and the tight TTFT SLOs they must satisfy. Hence, o

% \rui{Need to mention the requirement R1}

% \dmi{need to mention the online profiler here}

Given the prefillers' upstream location in the PD disaggregation system, the system must scale prefillers as soon as it detects an incoming burst to avoid violating \textbf{R1}. Hence, we introduce \emph{Convertible Decoders}, which are a subset of decoders that the system can quickly convert into prefillers upon a burst, and give back to the pool of decoders after the load spike on prefillers is over. Effectively, converting a decoder into a prefiller takes less than a millisecond to update the gateway routing rules to redirect the burst part of the request to the Convertible Decoder, because prefillers and decoders share the same model weights.
% Since resource demand increases almost immediately upon the arrival of request or token bursts, the policy that directly scales the prefiller, such as Blitzscale~\cite{zhang:blitzscale}, can cause severe TTFT violations. \textit{Convertible Decoder} is proposed to mitigate this, which can temporarily function as a prefiller. This design is motivated by two key observations: (1) the prefiller and decoder share identical model weights, and (2) the prefilling phase consumes relatively little GPU memory, while the decoding phase requires only modest GPU computation. As a result, the decoder’s spare resources can be repurposed to support prefilling tasks, provided their use is carefully managed to avoid SLO violations during decoding. 
Specifically, when a burst request arrives, the controller redirects the burst part of the request to the Convertible Decoder.
% \dmi{the below sounds like we just use the vanilla chunk-prefill technique, which is not true. Instead, say that we use an SLO-aware restricted chunk-prefill (see the subsec title) to guarantee high TTFT and TPOT SLO attainment }
In the Convertible Decoder, the prefilling phase employs an SLO-aware restricted prefill mechanism to maintain high TTFT and TPOT SLO attainment. After prefilling is finished, the same instance seamlessly continues with the decoding phase.

% the prefilling phase is executed with chunked prefill, where the chunk size is determined offline to ensure decoding SLOs are met~\cite{amey:sarathi}. 

% \dmi{seems to be an orphaned sentence, check}

% \rui{Add a brief paragraph about online profiling and SLO-aware chunked prefill.}
To ensure that Convertible Decoders effectively absorb bursty prefill requests without degrading the performance of active decoding tasks, \sys implements constraints on them. First, we regulate the prefill chunk size and allocate profile guided GPU memory reservations to maintain a stable token velocity that respects established SLOs. Furthermore, to avoid resource overprovisioning, an online profiling strategy dynamically determines the maximum number of convertible instances by monitoring peak burst sizes in traffic. By applying both restrictions, \sys limits the impact of sudden surges in traffic demand on the strict SLO of both prefill and decode requests. These bounds are validated empirically (\S\ref{sec:adapt}). 
\section{\sys Design}
In this section, we introduce \sys to address the different scaling requirements of the prefill and decode phases with the key design described in \S\ref{sec:core_method}.

\begin{figure}[t]
  \centering
  \includegraphics[width=\linewidth]{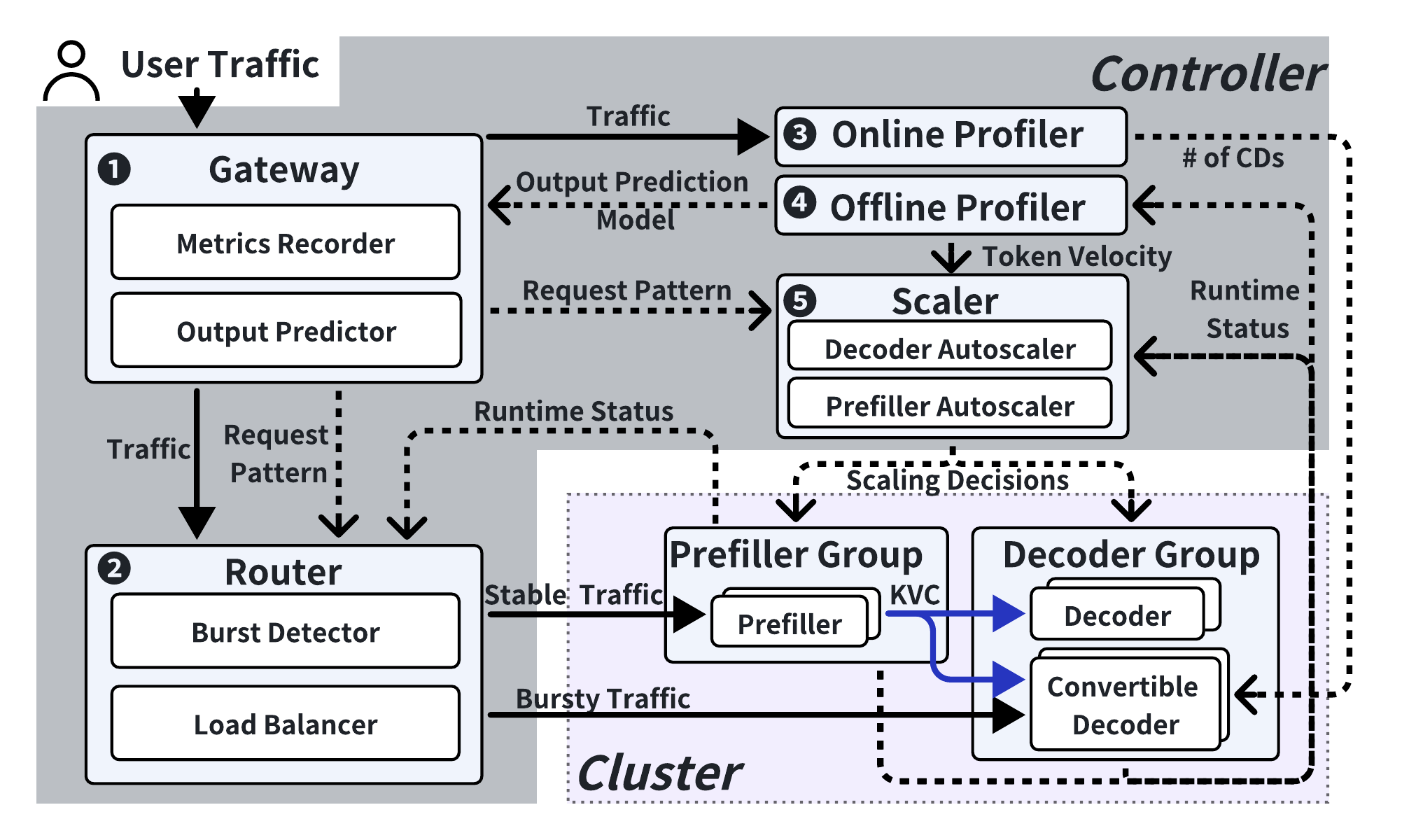}
  \caption{\sys architecture overview.
  \vspace{-10pt}
  }
  \label{fig:sys_architecture}
\end{figure}
\subsection{\sys System Architecture}
% \dmi{You need to start with a lead-in sentence: \sys introduces a novel controller that manages prefiller and decoder instances across the cluster.}
\sys introduces a controller comprising several components, shown in Fig.~\ref{fig:sys_architecture}, that manage prefiller and decoder instances across the PD disaggregated deployment cluster.
% As shown in Fig.~\ref{fig:sys_architecture}, the \sys controller comprises four main modules to serve LLMs within a Prefill/Decode~(PD) disaggregation system, namely \emph{Gateway}, \emph{Router}, \emph{Offline Profiler} and \emph{Scaler}. 
Upon receiving requests, the \emph{Gateway} (\ding{182}) records the number of incoming requests and their input token length, while also predicting the output length of each request with the output prediction model generated by the \emph{Offline Profiler} (\ding{185}).
The \emph{Online Profiler} (\ding{184}) continuously monitors arriving traffic over a fixed window, and uses the maximum burst size within this window to update the number of Convertible Decoders. Then the \emph{Router} (\ding{183}) employs a burst detector to distinguish burst traffic from stable traffic and routes them to appropriate instances based on request patterns with the \emph{Load Balancer}. Burst requests will be routed directly to the Convertible Decoders to finish their prefill and decode there, while stable requests are routed to regular prefiller and decoder in their respective groups. Meanwhile, the \emph{Scaler} (\ding{186}) monitors traffic patterns collected by the \emph{Gateway} and the runtime status of prefillers and decoders, and makes scaling decisions based on Token Velocity generated by the \emph{Offline Profiler}. 
% Since prefillers and decoders have different scaling requirements, the \emph{Scaler} incorporates dedicated Autoscalers for each.
Given the distinct scaling requirements, the \emph{Scaler} uses dedicated Autoscalers for prefillers and decoders.
% \dmi{and online profiler?}

% Meanwhile, the recorded request metrics collected by the \emph{Gateway} are also sent to the \emph{Decode Metrics Processor}(\ding{184}) and the \emph{Scaler}(\ding{185}) for generating the scaling policies of the prefiller and decoder. \emph{Decode Metrics Processor} employs an output predictor to predict the output length of each request and then classify it as different types according to the output length. Then the \emph{Scaler} can determine the number of the prefiller and decoder. The prefiller autoscaler collects the number of prefilling tokens and the status of the current prefiller to determine the number of prefillers, while the decoder autoscaler uses the classified metric from the \emph{Decoder Metrics Processor}, the request metric, and the current decoder status to determine the number of decoders and Convertible Decoders.

\subsection{Offline Profiler}
\label{sec:design_velocity}
% \TODO{this subsec is to be re-worked to follow the DynamoLLM style}

The key responsibility of the Offline Profiler is to construct models for the output predictor and to estimate the Token Velocity of the prefiller and decoder for the Autoscaler. 

\subsubsection{Output prediction model.}
Accurately estimating decoder Token Velocity requires predicting the length of each request. \sys adopts a lightweight output predictor similar to prior works~\cite{choukse:dynamollm, hu:deepserve}, which classifies requests based on input prompt content into different buckets by input-output lengths.
% , we use similar classification method to prior work~\cite{choukse:dynamollm}.
% \dmi{unclear what you mean by a bucket, be more specific. also, how many buckets and why (e.g., similar to prior work X)?}
% \cz{Provide the model parameters here.}
% Then \sys trains the classification model with an offline evaluation dataset. To further enhance its accuracy, \sys continuously collects ground-truth outputs from online inference and incrementally updates the model with these samples.

% \TODO{STOPPED HERE}

\subsubsection{Token velocity estimation.} 
% \TODO{This subsubsec duplicates \S\ref{sec:token_velocity}. The key is to move the conceptual part to \S\ref{sec:token_velocity} and here only focus on how the offline profiler calculates it. This sec has to shrink quite a lot.}
According to \S~\ref{sec:token_velocity}, Token Velocity refers to the maximum token release speed that the prefiller or decoder can achieve. This value is determined by both the model and the hardware characteristics; therefore, we perform offline profiling for each model-GPU pair to obtain its Token Velocity.

% \dmi{Remove $V_P$-s, we don't use them here. Use \emph{} and no capitalization for velocities}
To estimate the \emph{prefill velocity}, we gradually increase the request rate to the prefiller until its output rate saturates, which defines the maximum output velocity.

% \dmi{moved Vn to be in the middle}
For the \emph{network velocity}, we record the number of tokens transmitted from the prefiller to the decoder along with the corresponding transmission time. Based on these measurements, we compute network velocity as the maximum token transmission rate between the prefiller and the decoder.  In the following sections, we discount network velocity as it does not become the provisioning bottleneck in our evaluated RDMA setups, as we empirically demonstrate in \S\ref{sec:core_velocity_study}; it is still incorporated in the prefiller autoscaler to cover bandwidth-constrained deployments.
% \colin{Maybe you wanna cite some work here?}
% When multiple types of NICs are available, the Network Velocity should be estimated individually for each NIC. 

\emph{Decode velocity} is more complex to characterize because it varies across different types of requests. To derive this, we first overprovision the prefillers, sending request patterns, such as long input short output or short input medium output (see the full set of buckets in Tab.~\ref{tab:llama_decoder_tv} in \S\ref{sec:method}), and similarly sweep the request rate from low to high until the decoder reaches its peak output rate. The resulting maximum output rate for each request type is recorded as its decode velocity. 
We showcase the effectiveness of this method by evaluating its prediction vs. the optimal number of decoders for a production trace in \S\ref{sec:utilization}. 
% Notably, the decoder's Token Velocity differs from its maximum throughput: the latter reflects the rate at which new tokens are generated, whereas Token Velocity reflects the rate at which token generation is completed for entire requests.

\subsection{Scaling Decoders and Prefillers at Runtime}

% \subsubsection{Proactive Capacity Planning}
\sys's Scaler determines the number of prefillers and decoders from the ratio between the current token arrival rate and their Token Velocities. It then allocates the minimum instances required to provide sufficient Token Velocity for the incoming traffic. By projecting resource demand at traffic arrival, this strategy leaves enough reaction time for decoders downstream in the PD disaggregation system.

\subsubsection{Prefiller autoscaler.}
\label{sec:design_prefiller_as}
% \rui{Add prefix caching as a factor when calculating prefiller velocity}
% TokenPipe uses distinct scaling policies for each resource.
% \cz{remove network velocity?}
For prefillers, the required number of instances can be denoted as:

\begin{equation}
\small
I^P = \frac{\lambda \times (1-r_{kvc})}{min(V_P, V_{BW})},
\end{equation}
where $\lambda$ is the input token arrival rate, $r_{kvc}$ is the average KVC hit ratio.  $\lambda \times (1-r_{kvc})$ denotes the actual tokens that need to be processed by prefillers during specific time window. $V_P$ and $V_{N}$ are the prefiller and network velocities.
Once the new $I^P$ is different from the current instance number, the prefiller autoscaler triggers the autoscaling of the prefiller instance to boot up new instances or stop current running instances.
\subsubsection{Decoder autoscaler.}
\label{sec:design_decode_as}
% \dmi{need to open this subsec with: \sys manages two types of decoders: regular and convertible.}
\sys manages two types of decoders: regular decoders and Convertible Decoders, and calculates the total number of decoders as:
% \sout{the sum of the per-bucket decoder numbers required}
{\setlength{\abovedisplayskip}{2pt}
\setlength{\belowdisplayskip}{2pt}
\begin{equation}
\small
     I^D = \sum_{b \in \mathcal{B}} I^{(b)} = \sum_{b \in \mathcal{B}} \frac{\lambda'^{(b)}}{V_D^{b}}
\end{equation}
}
where $I^{(b)}$ denotes the decoder number required for bucket $b$, $\lambda'^{(b)}$ denotes the current input and output token arrival rate for request bucket $b$, $V_D^{b}$ denotes the profiled Token Velocity of bucket $b$, and $\mathcal{B}$ represents the set of all buckets. 
The autoscaler triggers scaling when the newly computed $I^D$ differs from the current number of decoder instances, following the same principle as for prefillers.

\subsection{Restrictions on Convertible Decoder}
% \rui{Add a subsection describe how to determine maximum number of Convertible Decoders.}
% \cz{Convertible Decoder uses capital word?}
\sys uses the Convertible Decoder to handle bursty prefill tasks by using chunked prefill in a SLO-aware restricted prefill strategy that we design.
% \dmi{re-phrase above highlighting that we introduce a novel algorithm here, not just "careful usage" or w/e it can mean}
To avoid interference with prefill and decode, we perform SLO-aware restricted prefill in Convertible Decoders and limit the maximum number of Convertible Decoders based on the maximum burst ratio over the past hour. 
\subsubsection{SLO-aware restricted prefill in Convertible Decoders.}
\label{sec:convertible_decoder}
While Convertible Decoders are designed to absorb traffic bursts, their configuration requires a careful balance to protect the SLOs of both incoming prefill tasks and ongoing decode tasks. The critical performance determinant for these instances is the \emph{chunk size}, which is defined as the maximum allowable sum of prefill tokens and the decode batch size within a single iteration. Selecting an optimal chunk size is essential; if the value is too large, the per-iteration latency increases significantly, which can lead to TPOT SLO violations. Conversely, choosing a small chunk size slows down prefill tasks. Thus, the chunk size must be sufficient to ensure that prefill tasks are processed without exceeding the TTFT SLO. 

For each model and hardware configuration, we profile the Convertible Decoder by gradually increasing the chunk size and selecting the largest one that still meets the TPOT SLO. Unlike the conventional chunked prefill method~\mbox{\cite{amey:sarathi}} that aims to reduce resource contention, our method maximizes prefill efficiency while keeping the TPOT within its SLO target.

Within each iteration, Token Velocity for prefill tasks in a Convertible Decoder is defined by the difference between the chunk size and the decode batch size as:
% \sout{. Thus, the prefill velocity of the Convertible Decoder is estimated as:}
% \dmi{what is TPOT exactly? TPOT SLO? what is BS?}
{\setlength{\abovedisplayskip}{3pt}
\setlength{\belowdisplayskip}{3pt}
\begin{equation}
\small
    V_D^{P'} = \frac{\mbox{chunk\_size} - \mbox{batch\_size}}{\mbox{TPOT}_{SLO}},
    \label{eq:conv_dec_tv}
\end{equation}
}
where the average decode batch size $\mbox{batch\_size}$ is estimated using available GPU memory divided by the average memory usage of the request in the trace, and $\mbox{TPOT}_{SLO}$ is the TPOT SLO requirement. Note that Convertible Decoders process no more than one prefill task at a time, returning to the decoder mode upon the prefill task's completion.
% \sout{if no more excessive prefill tasks occur.}
Therefore, \sys reserves a portion of GPU memory for prefill tasks, which is calculated as:
{\setlength{\abovedisplayskip}{3pt}
\setlength{\belowdisplayskip}{3pt}
\begin{equation}
\small
    Mem_{\mbox{$Reserved$}} = V_D^{P'} \times Mem_T \times \mbox{TTFT}_{SLO},
\end{equation}
}
where $V_D^{P'}$ represents the  Token Velocity of the Convertible Decoder and $Mem_T$ is the memory footprint of each token. 
\subsubsection{Number of Convertible Decoders.}
\label{sec:online_profiler}

% \dmi{the subsec title (and the text below) is misleading: online profiler is used to derive the number of CDs, not the max number (it sounds like there could be fewer CDs than the max, how is it derived?)}

% \rui{Add a section that describes the procedure of online profiling.}
% \begin{figure}
%     \centering
%     \includegraphics[width=\linewidth]{figures/design/online_profile.pdf}
%     \caption{Illustration of the maximum burst size over Azure code trace. The maximum number of Convertible Decoders is determined by the maximum burst size over the past hour.}
%     \label{fig:online_profile}
% \end{figure}
% The maximum number of Convertible Decoders is determined through online profiling...
The number of convertible decoders is determined through \emph{online} profiling based on the burst characteristics of a fixed-size sliding window. 
% \dmi{why one hour? this is a sensitivity parameter that we study}
To ensure the system can absorb the largest anticipated surge, it monitors the traffic across two time scales: \emph{instant traffic} and a per-minute window \emph{running average}.
% \dmi{I don't see anything about per-second traffic in 2.3}

Following the methodology in~\S\ref{sec:back_burst_wld}, we identify the maximum burst size as: {\setlength{\abovedisplayskip}{0pt}
\setlength{\belowdisplayskip}{2pt}
\begin{equation}
\small
    B = \max_{i\in t} \left( TPS(i) - TPS^t_{\text{avg}} \right),
\end{equation}}
where the $t$ is the sliding window size of the traffic, the $TPS(i)$ represents the token per second of each time point during the sliding window, and the $TPS^t_{\text{avg}}$ denotes the average TPS during the sliding window.
% Following the methodology in ~\S\ref{sec:back_burst_wld}, we identify the maximum burst size as the point where the instant traffic deviates most significantly from the running average \dmi{across a time window of GIVE DETAILS}. This maximum burst size $B$ is defined as the peak difference between these two metrics, measured in tokens per second:
% \begin{equation}
%     B = \max_{t} \left( TPS_{\text{inst}}(t) - TPS_{\text{avg}}(t) \right),
% \end{equation}
% where the $t$ is the sliding window size of the traffic, the $TPS_\text{inst}(t)$ and $TPS_{\text{avg}}(t)$ represents the token per second of the instant traffic $t$ and the 
The online profiler determines and periodically adjusts the number of convertible decoders $I^D_{r}$ as the ratio between the maximum burst size and the calculated prefill velocity of the Convertible Decoder in Eq.~\ref{eq:conv_dec_tv} as follows:
{\setlength{\abovedisplayskip}{-6pt}
\setlength{\belowdisplayskip}{-2pt}
\begin{equation}
\small
    I_r^{D} = \frac{B}{V^{P'}_D}.
\end{equation}
}
%  We first estimate the total number of decoders based on the maximum memory usage of a trace and the burst ratio of the trace using the same way as defined in~\S\ref{sec:back_burst_wld}. We then determine the number of Convertible Decoders $I^D_{c}$ by multiplying the estimated maximum number of decoders and the trace burst ratio. 
% Since Convertible Decoders must always be available to handle traffic bursts, \sys does not scale their number dynamically. 
% Instead, the autoscaler computes the required number of regular decoder instances as:
% \begin{equation}
% I^D_{r} = max(I^D - I^D_{c}, 0).
% \end{equation}

% \dmi{STOPPED HERE}
\subsection{Load Balancing Policy}
\label{sec:design_lb}
\setlength{\textfloatsep}{1pt}
\setlength{\floatsep}{1pt}
\captionsetup[algorithm]{skip=1pt}
\begin{algorithm}[t]
\caption{Routing Policy for Prefiller in \sys}
\label{alg:routing_prefiller}
\begin{algorithmic}[1]
\REQUIRE Prefill request $r$
\FOR{each prefiller $p$}
    \IF{$\frac{\textit{inflight\_tokens}(p)}{V_p}\le SLO(r)$}
        \STATE Assign $r \rightarrow p$; \textbf{return}
    \ENDIF
\ENDFOR
\FOR{each Convertible Decoder $d$}
    \IF{$\frac{\textit{inflight\_tokens}(d)}{V_D^{P'}} \le SLO(r)$}
        \STATE Assign $r \rightarrow d$; \textbf{return}
    \ENDIF
\ENDFOR
\STATE Enqueue $r$ to wait for an available prefiller
\end{algorithmic}
\end{algorithm}
% {\setlength{\textfloatsep}{3pt}
% \setlength{\floatsep}{3pt}
% \setlength{\intextsep}{3pt}
% \begin{algorithm}[t]
% \caption{Routing Policy for Prefill Requests in \sys}
% \label{alg:routing_prefiller}
% \small
% \begin{algorithmic}[1]
% \Require Prefill request $r$
% \For{each worker $w$ in prefiller-first order}
%     \State $V \gets \begin{cases}
%     V_p, & w \text{ is a prefiller}\\
%     V_D^{P'}, & w \text{ is a convertible decoder}
%     \end{cases}$
%     \If{$\frac{\textit{inflight\_tokens}(w)}{V} \le SLO(r)$}
%         \State Assign $r \rightarrow w$; \Return
%     \EndIf
% \EndFor
% \State Enqueue $r$
% \end{algorithmic}
% \end{algorithm}
% }

% \dmi{let's have two subsec: decod and prefill}
As an on-demand scaling system, \sys can create and terminate instances at any time, resulting in heterogeneous resource utilization across instances. Therefore, the load balancing policy is crucial to avoid SLO violations.

\subsubsection{Prefill load balancing policy.}
Although TTFT SLOs are tight, prefill task completion time is highly predictable because it is proportional to the input prompt length. Hence, we adopt a two-round strategy described in Alg.~\ref{alg:routing_prefiller}.
In the first round, the router checks whether any prefiller can satisfy the TTFT SLO. The waiting time is estimated as the ratio of in-flight tokens to the prefiller’s Token Velocity. If no prefiller can accommodate the request, the router enters the second round and iterates over all Convertible Decoders. For each convertible decoder, it compares the in-flight tokens with its prefill Token Velocity to determine whether it can serve the incoming request within the TTFT SLO. If neither round finds a feasible worker, the router enqueues the request until a prefiller becomes available. As system load changes, due to decoder utilization changes or the initialization of additional prefilling instances, \sys Scaler re-evaluates Convertible Decoders and re-assigns queued requests accordingly.
\subsubsection{Decode load balancing policy.}
For decode requests, TPOT violations occur either due to insufficient memory or because the maximum batch size is capped.
For each incoming decode request, the router first uses the output predictor to classify it into a request type (e.g., short input and long output). It then checks all decoders and routes the request to the decoder with the fewest in-flight requests of that type to balance the per-type load across decoders. Convertible decoders are excluded from routing once their memory utilization exceeds the predefined threshold. On Convertible Decoders, the router further prioritizes decode requests over prefill tasks to ensure that decoding is not preempted by prefill execution.

\subsection{Multi-turn Workloads and Session Affinity}
Production deployments may enforce session-affinity routing, e.g., to reuse the prefix KVC of multi-turn conversations on the same decoder. Such constraints shrink the pool of decoders eligible to act as Convertible Decoders, since a decoder that also serves affinity-bound decode requests has less residual capacity for prefill. This work focuses on single-turn workloads; we leave extending \sys to multi-turn workloads with session affinity to future work.

\subsection{\sys Implementation}
% \dmi{Let's shorten it a lot, dropping component enumeration and unnecessary details: We build \sys control plane with 6000 Golang LoC. We integrate metrics reporting with Prometheus to retain compatibility with vLLM and Kubernetes metric monitoring subsystems. that's it}
% \dmi{Might need to add impl details on the challenging parts we faced: LMCache integration, esp with TP took us some time to figure out; handling multi-node clusters and low-latency / pipelined network transfers; anything else?}

% \cz{Does the offline profiler use Go? Better use less 'we'}
% We use vLLM to implement the PD disaggregation framework with LMCache as the KV-cache transfer backend.
\sys is built on top of vLLM~\mbox{\cite{kwon:efficient}}v0.9.2, using LMCache~\mbox{\cite{liu:lmcache}} as the KVC transfer backend to support PD disaggregation inference. We support asynchronous KVC transfer in LMCache by using dedicated I/O threads on the prefiller and decoder to perform send and receive operations independent of the main computation. This design, built on the NIXL communication backend, allows KVC movement to proceed in parallel with computation and prevents additional latency. We build \sys control plane with 6000 Golang LoC. We integrate metrics reporting with Prometheus to retain compatibility with vLLM monitoring subsystems.
The decoder-to-prefiller conversion is implemented by routing prefill tasks to the Convertible Decoder at a higher priority, so vLLM co-schedules them with ongoing decodes under the SLO-aware chunk size. 

\section{Experimental Methodology}
\label{sec:method}

% \dmi{RevA asked for clarification of the evaluated scenario: need to say that we model that the provider has a fixed-size pool of nodes equipped with GPUs and allocate them on demand to the served workload, leaving unallocated GPUs and nodes stand-by, or potentially allocated to other workloads (not modeled).}

% \subsection{Experimental Setup}
\label{sec:setup}
\noindent\textbf{Hardware setup.}
To demonstrate \sys's performance on different hardware, we evaluate \sys on two GPU clusters: AMD cluster and Nvidia cluster. The AMD cluster is equipped with 8 MI308X nodes; each node has 8 MI308X GPUs with four 200 Gb/s RDMA links through Mellanox BlueField-3, providing 800 Gb/s aggregate inter-node RDMA bandwidth.
The Nvidia cluster consists of 16 nodes, each equipped with 4 A100-40GB GPUs with two Mellanox ConnectX-6 NICs (total 200 Gbps RDMA).
In our evaluation, we model a scenario where the provider maintains a fixed-size pool of GPU nodes and allocates them on demand to the served workload. Unallocated GPUs and nodes remain on standby or may be allocated to other workloads (out of our scope).

% The H100 cluster has two nodes, each with 8 H100-80GB GPUs with NVLink (total1200 GB/s) and twelve Mellanox ConnectX-6 NICs (total 2880 Gbps RDMA).
% \dmi{typically Mellanox uses Infiniband, not ROCE; can you double check?}
% \dmi{NIC BW should be in Gb/s not GB/s (historical reasons)}

% \dmi{Software setup? e.g., vLLM version, CUDA version, etc.}
% \dmi{RDMA NIC versions and BW; NVLink BW also}

% \noindent\textbf{System Configuration}
% \cz{Please write something here}
\noindent\textbf{Software setup.}
We deploy \sys on top of vLLM~\cite{kwon:efficient} v0.9.2, using LMCache~\cite{liu:lmcache} v0.3.0 as the KVC transfer backend. The evaluation runs on PyTorch v2.7.1 and CUDA v12.8.
We adopt the ServerlessLLM model loader~\cite{fu:serverlessllm} to minimize the vLLM and model initialization delays. For the output predictor, we finetune a 110M BERT-base model~\cite{qiu2024efficientinteractivellmserving} on 8 AMD MI308X GPUs on ShareGPT trace. The model classifies each request's output length bucket from its input prompt. For traces with actual prompt content, such as ShareGPT~\cite{sharegpt}, we directly apply this output predictor. For traces without prompt content, such as the Azure Conversation and Azure Code traces~\cite{choukse:dynamollm}, we set the output predictor at the same accuracy as measured on the ShareGPT trace. The accuracy and overhead of the predictor are evaluated in \S\ref{sec:eval_accuracy}. For the online profiler, we use a 1-hour window to accurately estimate the required Convertible Decoder capacity. The impact of window size is evaluated in \S\ref{sec:online_profiler_acc}.

\noindent\textbf{Models setup.}
In our evaluation, we use two models: a dense model \llama, and a MoE model \Qwen. Both models are served in half precision (bfloat16). We deploy the \llama with a tensor parallelism degree of one (TP=1) on the Nvidia cluster. We deploy the \Qwen with a tensor parallelism degree of four (TP=4) on the Nvidia cluster and a tensor parallelism degree of one (TP=1) on the AMD cluster. Following \sllm~\cite{fu:serverlessllm}, we cache the model weight on each node's CPU memory to achieve low initialization latency (about 5s for \llama and 8s for \Qwen).

\noindent\textbf{Baselines.}
We compare \sys against four representative baselines under the PD disaggregation: 
% \dmi{specify how we constructed the baselines (e.g., used the original code of their specific algo, re-implemented atop our prototype carefully modeling X and Y to isolate the implications of the autoscaling control plane (smth like that, must sound reasonable and robust)}
We re-implement the autoscaling policies atop our vLLM-based prototype, carefully controlling for autoscaling sampling frequency, traffic metrics smoothing, and instance cold-start latency, to isolate the implications of the autoscaling control plane from other optimization and configuration differences in the runtime versions the authors used. 
\begin{itemize}[leftmargin=0cm,itemindent=.4cm,labelwidth=\itemindent,labelsep=0cm,align=left, partopsep=0pt,itemsep=2pt,parsep=0pt]
    \item \emph{\aibrix} uses a request-based autoscaler for prefillers, monitoring concurrency, and a utilization-based autoscaler for decoders, tracking GPU memory usage. The prefiller threshold is the ratio of maximum prefill throughput to the trace’s average prefill length, while the decoder threshold is $70\%$ GPU memory.
    \item \emph{\blitz} employs a request-based autoscaler for both prefillers and decoders and implements \emph{live autoscaling}, which enables prefillers to start KVC computation during model loading. We emulate ideal live autoscaling by recording scaling decisions and executing scale-up actions proactively, effectively removing model loading latency from the critical path. Prefiller thresholds use the ratio of the maximum prefill throughput to the trace’s average prefill length, decoder thresholds are based on the ratio between available KVC memory and the average memory footprint of requests.
    \item \emph{\dist}. DistServe scales both prefillers and decoders using a simulator that accounts for the input request rate, the profiled model, and the hardware characteristics to maximize goodput.
    \item \emph{\spw} uses the same scaling policy and threshold determination as \dist, but with mixed pool enabled, where prefillers and decoders can be converted into instances with mixed continuous batching to absorb traffic bursts in the corresponding stage. 
\end{itemize}
We provide an example of the scaling thresholds of all evaluated systems for Llama-3.1-8B on the A100 cluster in Table~\ref{tab:llama_scaling}. The decoder thresholds in \sys are not directly applicable, because they are dynamically determined by per-bucket Token Velocity, which varies with the numbers of input and output tokens. As described in \S\ref{sec:design_velocity}, we derive the per-bucket Token Velocity by sweeping the request rate until token output rate saturates. We provide an example for Llama-3.1-8B in Table~\ref{tab:llama_decoder_tv}.

\begin{table}[t]
\centering
\small
\setlength{\tabcolsep}{3pt}
\begin{tabular}{llcccccc}
\toprule
System & Threshold unit
& \multicolumn{2}{c}{Azure Conv.}
& \multicolumn{2}{c}{Azure Code}
& \multicolumn{2}{c}{Mixed} \\
\cmidrule(lr){3-4} \cmidrule(lr){5-6} \cmidrule(lr){7-8}
& & P & D & P & D & P & D \\
\midrule
AIBrix & P: req, D: \% & 7 & 70\% & 7 & 70\% & 11 & 70\% \\
BlitzScale & P: req, D: req & 7 & 45 & 7 & 38 & 11 & 41 \\
DistServe & req/s & 14 & 28 & 8 & 20 & 12 & 23 \\
SplitWise & req/s & 14 & 28 & 8 & 20 & 12 & 23 \\
\sys & P: tok/s, D: N/A & 14K & N/A & 14K & N/A & 14K & N/A \\
\bottomrule
\end{tabular}
\caption{Scaling thresholds for Llama-3.1-8B (TP=1) on A100 under different traces.}
\label{tab:llama_scaling}
\end{table}

\begin{table}[t]
\centering
\small
\setlength{\tabcolsep}{2pt}
\begin{tabular}{lccccccccc}
\toprule
& S--S & S--M & S--L & M--S & M--M & M--L & L--S & L--M & L--L \\
\midrule
Input & 256 & 256 & 256 & 1K & 1K & 1K & 8K & 8K & 8K \\
Output & 100 & 350 & 610 & 100 & 350 & 610 & 100 & 350 & 610 \\
Tok. Vel. & 23,535 & 8,146 & 5,138 & 33,106 & 9,794 & 5,766 & 39,551 & 11,310 & 6,495 \\
\bottomrule
\end{tabular}
\caption{Per-bucket decoder Token Velocity (Tok. Vel.) for Llama-3.1-8B (TP=1) on A100.}
\label{tab:llama_decoder_tv}
\end{table}

\noindent\textbf{Workload Generation.}
We evaluate \sys using traces from the Azure~\cite{choukse:dynamollm}, BurstGPT~\cite{wang2025burstgpt} and ShareGPT~\cite{sharegpt}. 
To examine performance under different workloads, we construct three representative traces: 
\emph{Azure Conversation} (conv), \emph{Azure Code} (code), and a \emph{Mixed} trace (mix) combining all Azure Conversation, Azure Code, and BurstGPT traces with equal request rates. Since ShareGPT does not contain request arrival patterns, we use the Azure Conversation trace's timestamps to send ShareGPT requests.
We use a custom load generator to replay these traces while preserving their input/output token distributions and request arrival patterns. As the Azure and BurstGPT traces provide arrival times and input/output lengths but no session identifiers or prompts, requests are replayed independently
Following~\cite{choukse:dynamollm}, where the Azure trace was collected on an 11-node H100 cluster serving Llama-2-70B, we scale the Azure Conversation trace according to the ratio between the maximum memory capacity of our setup and that of their deployment. 
The sampled trace yields an average throughput of 22~RPS. 
We then recursively downsample Azure Code and Mixed to the same RPS for consistent comparison across workloads. 
Because Azure traces, BurstGPT and ShareGPT contain at most 8K input tokens, we additionally evaluate \sys under long-context workloads by scaling request input lengths by 4$\times$ when serving \Qwen.
% \rui{To evaluate performance under long-context workloads, we scale the input length to 32K for \Qwen.} \cz{\Qwen supports 40K context length, and the SLO only sets for 8k. Remove it.}
% \rui{Long input length workload.}
% \dmi{replace with a shorter note somewhere above: (token length scaled up to 32K for \Qwen)}
% The maximum input length for both Azure and BurstGPT is only 8K. To evaluate \sys's performance under long-context workloads, we scale the input request length by 4x when serving \Qwen models.

\noindent\textbf{Metrics.}
% \cz{how can we calculate the SLO attainment? each token has one TPOT, do we use average or max? }
We use SLO attainment rate to evaluate an autoscaling system's performance and the average utilized GPU numbers during trace run to measure its cost.
We follow the SLO standards defined in prior works~\cite{choukse:dynamollm,mlperf}, which assign different TTFT targets based on input length: 250~ms for short requests ($<256$ tokens), 400~ms for medium requests ($<1024$ tokens), and 2000~ms for long requests (up to 8192 tokens). The TPOT SLO is fixed at 100~ms. 
% \dmi{define the perf/cost metric}

\begin{figure}[t]
  \centering
  \begin{subfigure}[t]{\linewidth}
    \includegraphics[width=\linewidth]{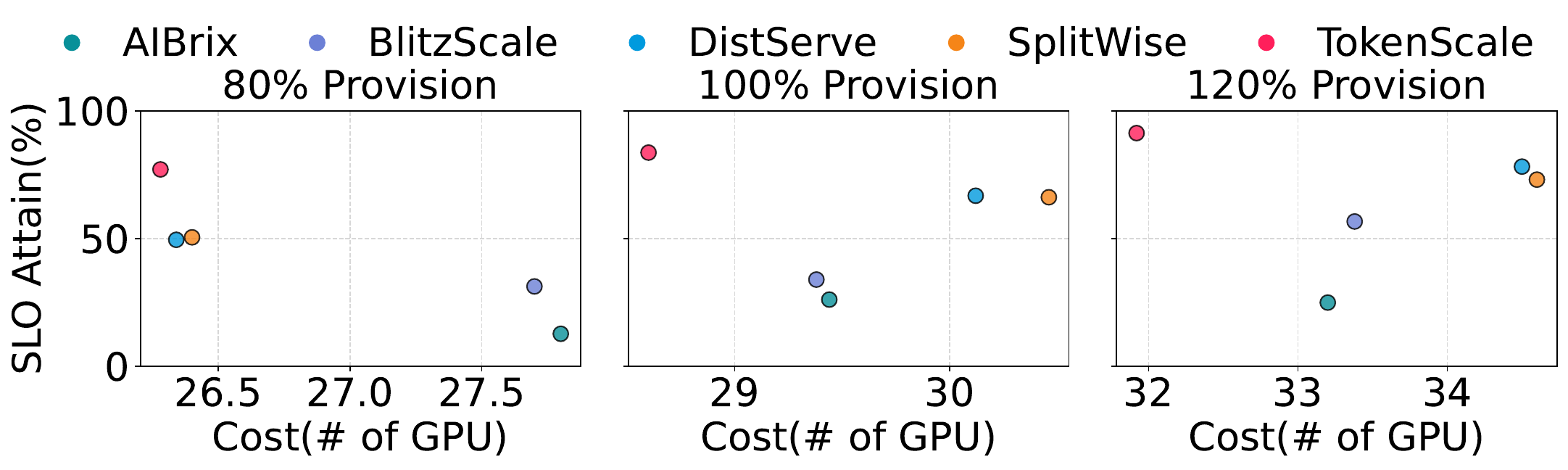}
    \caption{Performance vs. cost. \emph{\textbf{Top-left is better.}}}
    \label{fig:qwen_amd_e2e}
  \end{subfigure}
    \begin{subfigure}[t]{\linewidth}
    \includegraphics[width=\linewidth]{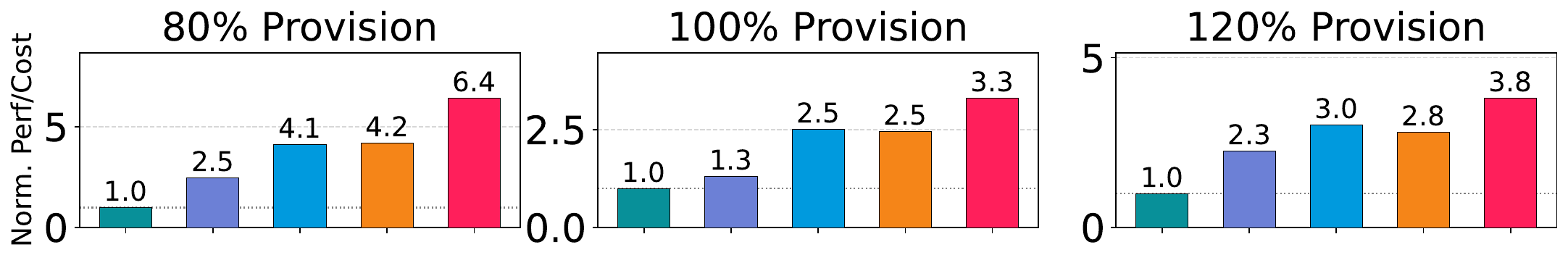}
    \caption{Normalized performance/cost.\vspace{-10pt} }
    \label{fig:qwen_amd_norm}
  \end{subfigure}
  \caption{Comparison of performance vs. cost of \Qwen model running ShareGPT traces on AMD cluster. We sweep the provisioning ratios to compare different systems' performance under $80\%$, $100\%$ and $120\%$ resource provisioning.}
\end{figure}

\begin{figure}[t]
  \centering
  \begin{subfigure}[t]{\linewidth}
    \includegraphics[width=\linewidth]{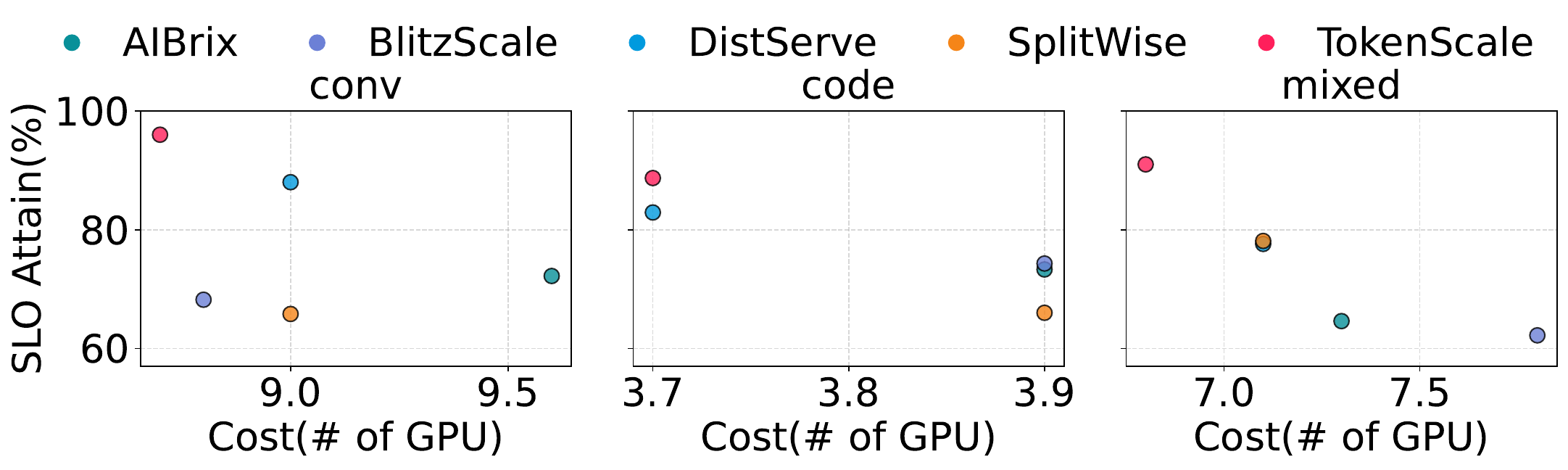}
    \caption{\llama}
    \label{fig:llama3_e2e}
  \end{subfigure}
  \hfill
  \begin{subfigure}[t]{\linewidth}
    \includegraphics[width=\linewidth]{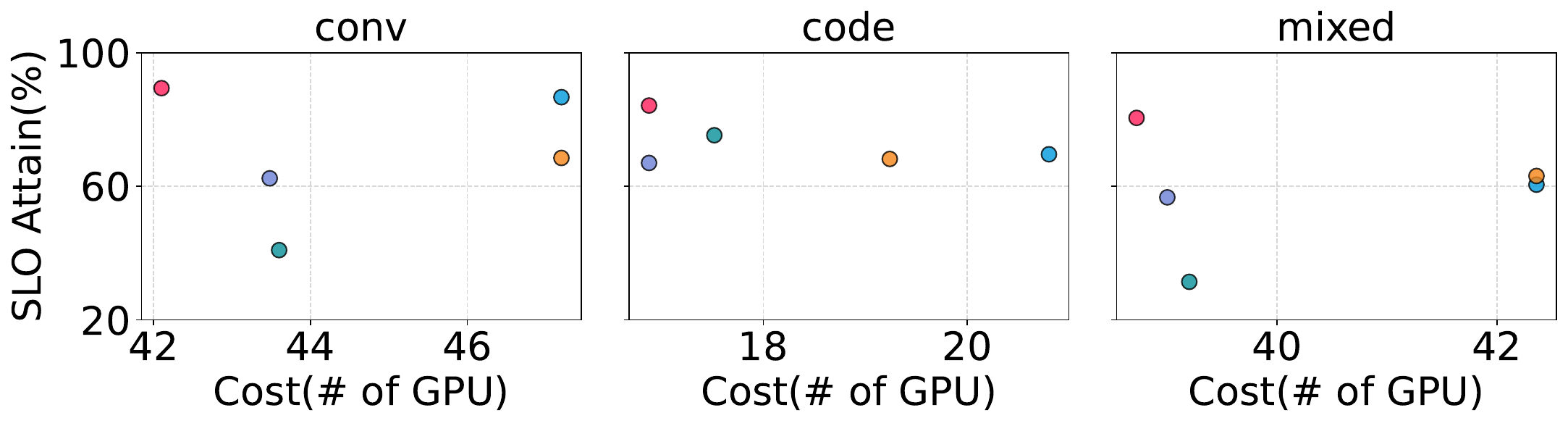}
    \caption{\Qwen\vspace{-10pt}}
    \label{fig:qwen_e2e}
  \end{subfigure}
  \caption{Comparison of performance vs. cost across different scaling systems, traces, model configurations: (a) \llama, TP=1; (b) \Qwen, TP=4 on the Nvidia cluster.
  \emph{\textbf{Top-left is better.}}}
  % \TODO{Fix up the legend; let's make these figs larger, they are too small and too important}
  % \colin{maybe for this figure, it may help to put an arrow to show which direction is better. Top left yea?}
  \label{fig:e2e}
\end{figure}
\section{Evaluation Results}

% \dmi{need a lead-in with 2-3 sentences, describing the structure of the eval sec.}
In this section, we evaluate the design and performance of \sys under various workload patterns and scaling scenarios.
We conduct two end-to-end experiments on AMD and Nvidia clusters. On the AMD cluster, we conduct a study comparing performance and cost under different resource provisioning levels of \sys and all baselines. On the Nvidia cluster, we further demonstrate \sys's performance across different models, traces and tensor parallelism configurations. Next, we analyze the effectiveness of key components in \sys, including burst adaptation, comparison of provisioned vs. required instance number, output predictor's impact on \sys, and online profiler's accuracy. We then conduct an ablation study to quantify the contribution of components to overall performance.

\subsection{End-to-End Experiments}
\label{sec:e2e}
% We conduct end-to-end experiments on AMD and Nvidia clusters: on AMD, we sweep the provisioning ratio. On Nvidia, we vary traces, models, and parallelism settings.

\subsubsection{End-to-end experiments on AMD cluster with different provisioning ratios.}
To compare different autoscaling policies under various resource provisioning levels, we apply a provisioning ratio to autoscalers. For example, a provisioning ratio of 80\% scales the autoscaler's instance count by 0.8. As shown in Fig.~\ref{fig:qwen_amd_e2e}, \sys consistently improves the SLO attainment rate by $26.6$-$64.4\%$, $16.9$-$57.6\%$, and $13.1$-$66.4\%$ under $80\%$, $100\%$, and $120\%$ resource provisioning, respectively, with $0.2$-$5.5\%$, $2.7$-$6.1\%$, and $3.9$-$7.8\%$ lower cost. Overall, \sys improves the SLO attainment rate by $13.1$-$66.4\%$ with $0.2$-$7.8\%$ lower cost. Fig.~\ref{fig:qwen_amd_norm} shows the normalized performance/cost, over all provisioning levels, \sys improves performance/cost by $1.3$-$6.4\times$. This demonstrates \sys's ability to deliver the right amount of resources to achieve SLO attainment across varying provisioning levels.

\subsubsection{End-to-end experiments on Nvidia cluster with different traces and models.}
Fig.~\ref{fig:e2e} demonstrates that \sys achieves higher SLO attainment with fewer GPUs than all baselines across different workloads and model sizes. As shown in Fig.~\ref{fig:llama3_e2e}, \sys improves SLO attainment from $62$-$88\%$ to $89$-$96\%$ across workloads while using $6$-$13\%$ fewer GPUs on average for the \llama. For the \Qwen (Fig.~\ref{fig:qwen_e2e}), even under long-context workloads, \sys improves SLO attainment from $32$-$87\%$ to $81$-$90\%$ while using $3$-$11\%$ fewer GPUs than the baselines.

This advantage arises because baseline systems rely on concurrency-based autoscalers for prefillers, which respond suboptimally to sudden traffic bursts and lead to noticeable fluctuations in prefiller instances. \sys benefits from two key design factors: 1) the Token Velocity-based scaling policy accurately estimates resource demand, enabling timely and balanced provisioning without over-/under-scaling; and 2) the Convertible Decoder, supported by accurate online profiling and dedicated memory reservation, can absorb bursty prefill traffic without degrading decoder throughput. 
In our experiments, a single Convertible Decoder, as determined by the online profiler, is sufficient to absorb bursts in the production traces we used.
% \rui{The online profiler determines that one Convertible Decoder suffices to absorb the maximum observed burst across all configurations.}
As for \spw, it fails to avoid prefill-decode interference, leading to performance degradation. Under \spw, most prefillers operate with chunked prefilling, preventing them from reaching optimal prefill throughput. Moreover, decoders without memory reservation cannot immediately convert when the decode memory is saturated.

% \dmi{specify how many (or \%) of GPUs allocated to conv decoders in our experiments (RevA)}

\subsection{Effectiveness of \sys Design}
\label{sec:effectiveness}

\subsubsection{Adaptation to burst workload.}
\label{sec:adapt}
% \dmi{this subsection is odd because it does not discuss or compares to any other baselines. This needs to be fixed}

% In this section we're going to analyze where does the benefit of Tokenpipe come from.
% \cz{How the autoscaling policies and Convertible Decoder adapt to bursty traffic .}
% F13,F14
% \dmi{furthermore is out of place here by its meaning}
We analyze how \sys adapts to burst workloads. 
% \cz{how many is Convertible Decoder used here?}
% Specifically
% \dmi{in this experiment}
In this experiment, we start the system from 1 prefiller and 1 Convertible Decoder to serve a stable traffic of 1 RPS. At time $t=10s$, the RPS is scaled to 10 to emulate a traffic burst.   
% \hong{not very clear that the throughput is generated tok/s or the Token Velocity, what is the req setting}
% We collect the TTFT vs. time around the spike and demonstrate it in Fig.~\ref{fig:adapt}.
\begin{figure}[t]
  \centering
  \begin{subfigure}[t]{\linewidth}
    \includegraphics[width=\linewidth]{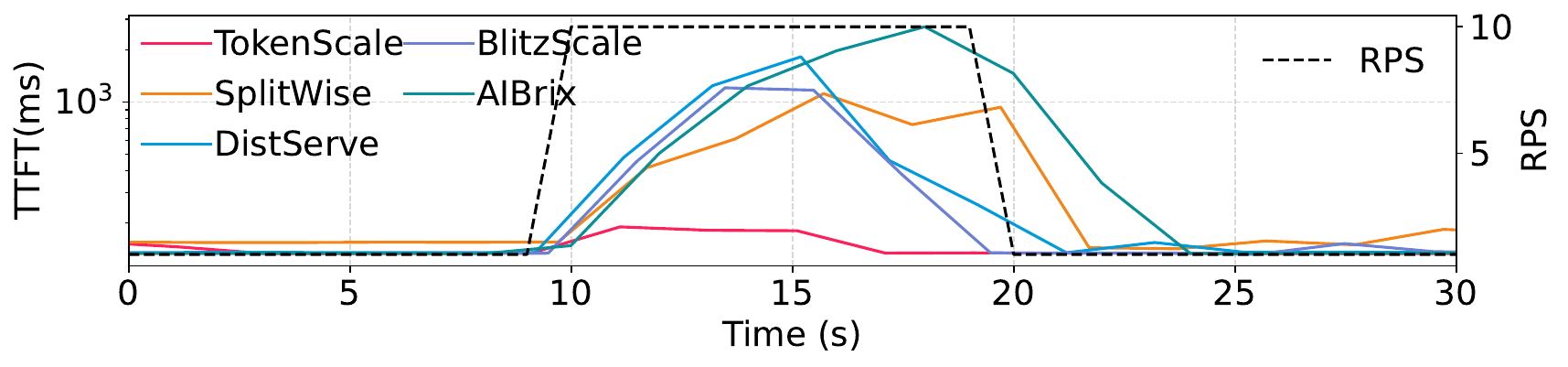}
    \caption{TTFT}
    \label{fig:adapt_ttft}
  \end{subfigure}
  \hfill
  \begin{subfigure}[t]{\linewidth}
    \includegraphics[width=\linewidth]{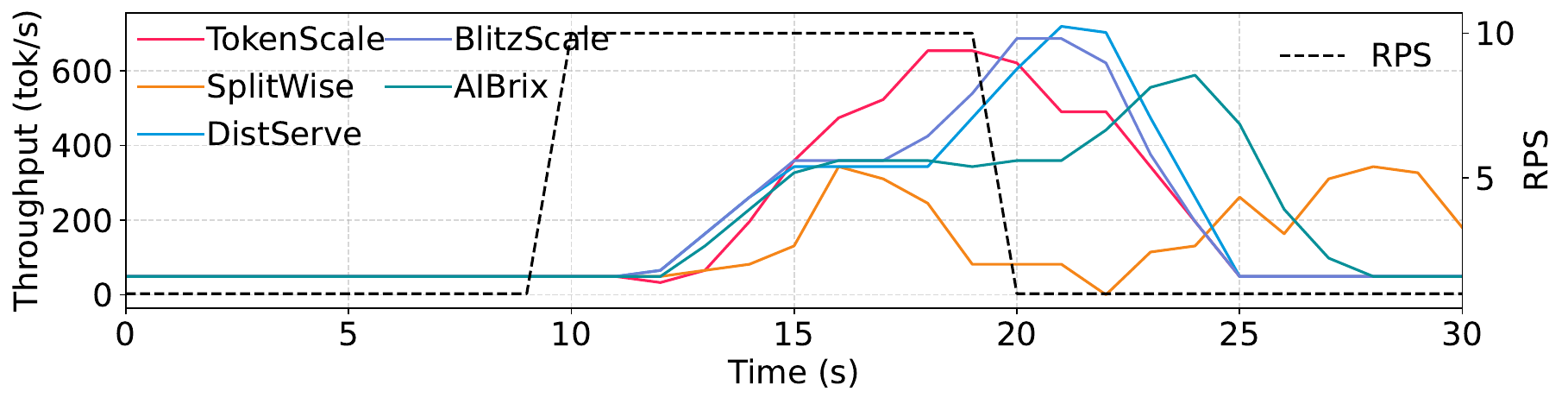}
    \caption{Decoder throughput\vspace{-10pt}}
    \label{fig:adapt_throughput}
  \end{subfigure}
  \caption{TTFT and decode throughput vs. time for different systems running the bursty trace in the presence of a $10\times$ burst at $t=10s$ for \llama.
  % \rui{Need to add texts to explain Splitwise}
  }
  \label{fig:adapt}
\end{figure}
% \begin{figure}
%     \centering
%     \includegraphics[width=\linewidth]{figures/tokenscale_outputs/adaptation.pdf}
%     \caption{Different system's ability to adapt to traffic burst.
%     % \cz{Require to show how the traffic burst looks like (on the secondary Y axis). Refer to Fig.4} \dmi{try log Y axis}
%     }
%     \label{fig:adapt}
% \end{figure}
% \cz{How many is the TTFTs of different systems? Please make it clear. It is better to first show the result and then explain why this result occurs. We need to directly show that \sys is better than the baselines, rather than distributing in different parts.}
Fig.~\ref{fig:adapt_ttft} shows that \sys experiences only a slight TTFT increase, rising to about $50$\,ms, and quickly recovers at $t=14$s. In contrast, all other baselines exhibit significant TTFT surges, reaching above 1000ms, and recover much later than \sys. 
\sys maintains low TTFT by redirecting bursty requests to the Convertible Decoder. 
% \rui{Rev C1's concern} 
We further evaluate the mechanism’s impact on Convertible Decoder throughput and observe only a drop of less than $10\%$ at $t=12$\,s (Fig.~\ref{fig:adapt_throughput}).
In contrast, while \spw also redirects bursty traffic to decoders, it lacks a dedicated memory reservation mechanism like the \sys Convertible Decoder. 
Consequently, this leads to severe prefill-decode interference, causing both higher TTFT and lower decoder throughput.
Overall, \sys reacts much faster to changes in traffic than the baselines, due to its Convertible Decoder and rapid prefiller scaling policy, while incurring only a negligible degradation in decode throughput.

\subsubsection{Provisioned vs. required number of instances.}
\begin{figure}
    \centering
    \begin{subfigure}[b]{\linewidth}
    \includegraphics[width=\linewidth]{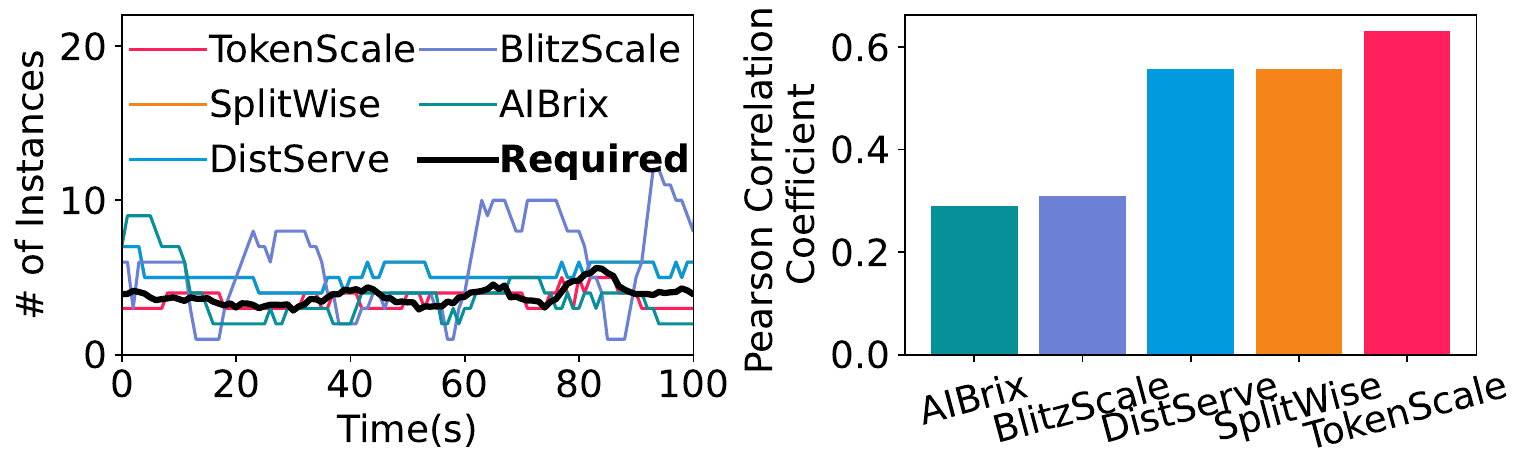}
    % \vspace{-10pt}
    \caption{Prefiller.} \label{fig:scaling_efficiency_prefill}
    \end{subfigure}
    \begin{subfigure}[b]{\linewidth}
    \includegraphics[width=\linewidth]{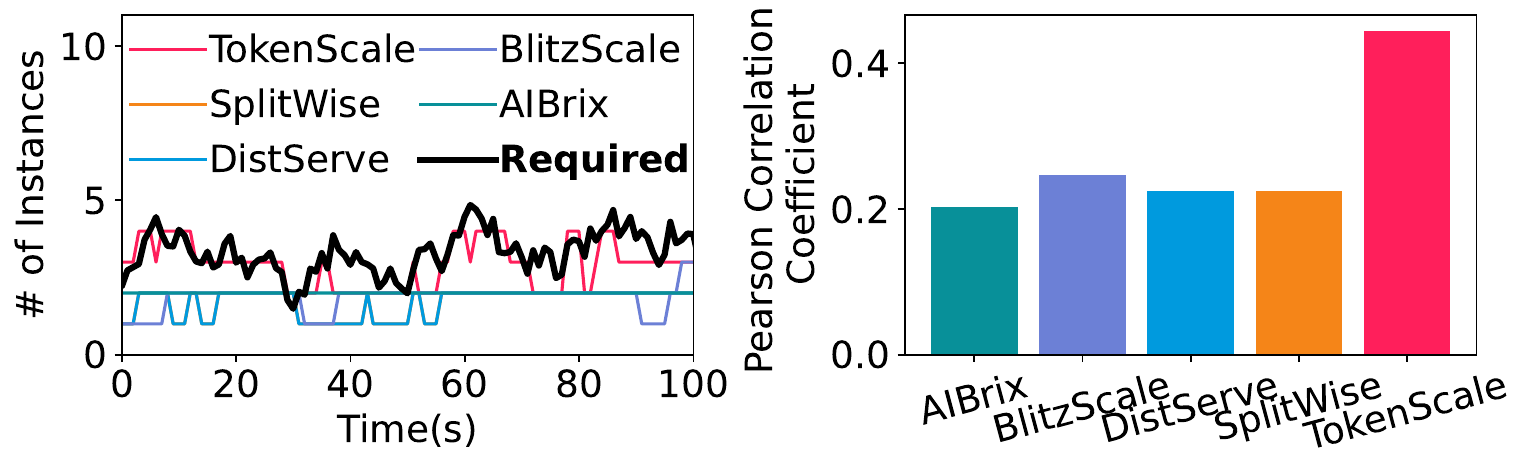}
    \caption{Decoder.\vspace{-10pt}}  
    \label{fig:scaling_efficiency_decode}
    \end{subfigure}
    \caption{
    % Left: Required number of instances (black line) compared to the instance numbers chosen by different autoscaling policies (colored). The black curve shows the ground-truth required instance number, while the colored curves represent the provisioned instance number chosen by different autoscaling policies. Right: Pearson correlation coefficients between the provisioned instance number and the required number. Note that as \spw uses the same scaling policy as \dist, thus they have the same trend and result.
    Left: Required instance number (black) versus the number of instances provisioned by different autoscaling policies (colored). The black curve denotes the ground-truth required number of instances, while the colored curves show the numbers provisioned by different autoscaling policies. Right: Pearson correlation coefficients between the provisioned and required instance numbers. Note that \spw uses the same scaling policy as \dist and therefore exhibits the same trend and results.
    }
    \label{fig:scaling_efficiency}
\end{figure}

\label{sec:utilization}
% \cz{Poor segmentation. Please re-organize this section.}
% In the previous subsection, we analyzed how \sys adapts to a single burst workload. Here, we examine 
We evaluate how different scaling policies provision instances under a trace.
We measure the correlation between the provisioned and required number of instances over time across different scaling strategies to assess the effectiveness of the proposed Token Velocity-based policy and mechanisms.
To determine the required number of instances, we provision sufficient GPUs for both the prefiller and decoder, and then derive the ground-truth instance requirements for each component by multiplying their measured resource utilization by the total GPU capacity allocated. Prefilling throughput and decoder memory occupancy are used as utilization metrics for prefillers and decoders, respectively.

% Specifically, the required instance number is obtained by multiplying the overprovisioned instance count by the measured utilization.
% For example, when running the trace with eight prefill instances, if the compute utilization at a certain timestamp is 20\%, the required instance number is computed as $8 \times 20\% = 1.6$. 
% \cz{You can directly introduce your experiment setting in the Fig caption.}
% In our experiment, we use the same workloads as in Fig.~\ref{fig:llama3_e2e}, and configure the overprovisioned cluster with 8 prefillers and 8 decoders. 
% \cz{DistServe should be mentioned as it is comparable.}

% \cz{rewrite this part, merge two paragraphs as one. The explanation of a figure should not stray too far from the figure's reference.}
% We plot the required and provisioned instance numbers as black and colored curves in Fig.~\ref{fig:scaling_efficiency_curve_prefill} and Fig.~\ref{fig:scaling_efficiency_curve}. 
% The Pearson correlation coefficient between the two is reported for each policy in Fig.~\ref{fig:scaling_efficiency_bar_prefill} and Fig.~\ref{fig:scaling_efficiency_bar}. 
Fig.~\ref{fig:scaling_efficiency} compares the number of instances each system provisions against the ground truth. For prefillers, \sys achieves the highest correlation coefficient ($0.63$) between allocated and required instance counts. \dist and \spw also correlate relatively well, as they derive scaling thresholds from an accurate LLM simulator, but their request-based policies lag behind workload dynamics. In contrast, \aibrix and BlitzScale fluctuate significantly, as their concurrency-based autoscalers react slowly to bursts and scale up only once requests queue. For decoders, \sys achieves a correlation coefficient of $0.44$, about twice that of the baselines, since its pre-profiled decoder velocities (Tab.~\ref{tab:llama_decoder_tv}) enable proactive and accurate scaling, whereas the baselines scale decoders with significant delay and inaccuracy.
 
% As a result, compared with \sys, baselines tend to overprovision when the burst arrives and then scale down aggressively after the burst subsides.
\subsubsection{Output predictor.} 
% \rui{Going to put a study to demonstrate the real accuracy of output predictor}
\label{sec:eval_accuracy}
In this section, we first evaluate the accuracy and overhead of \sys's output predictor, then study \sys's sensitivity to predictor accuracy.
We measure the predictor's accuracy as the fraction of prompts correctly classified into their corresponding output token buckets on the ShareGPT trace. The predictor achieves an accuracy of $60\%$ with an overhead of $5$~ms on the AMD cluster and $8$~ms on the Nvidia cluster, as shown in Fig.~\ref{fig:acc_overhead}. The overheads are both negligible compared to the $2000$~ms prefiller SLO.
To further study how predictor accuracy affects \sys's performance, we conduct a controlled experiment by sweeping the accuracy of the output predictor. We replace the original output predictor with a simulated one with varying accuracy from $100\%$ to $50\%$ and run the ShareGPT trace under each setting, measuring both SLO attainment and GPU usage. 
As shown in Fig.~\ref{fig:accuracy}, \sys maintains high SLO performance across all accuracy levels: when accuracy drops from $100\%$ to $50\%$, the SLO attainment rate decreases by merely $2\%$ while cost increases slightly by $6\%$. 
This demonstrates \sys's robustness to predictor accuracy since prediction errors cause only minor overprovisioning without degrading overall performance. Moreover, as the red dashed line shows our current predictor's accuracy $60\%$, these results highlight \sys's potential for further cost reduction as more accurate prediction methods become available.

\begin{figure}
    \centering
    \begin{subfigure}[!t]{0.49\linewidth}
        \includegraphics[width=\linewidth]{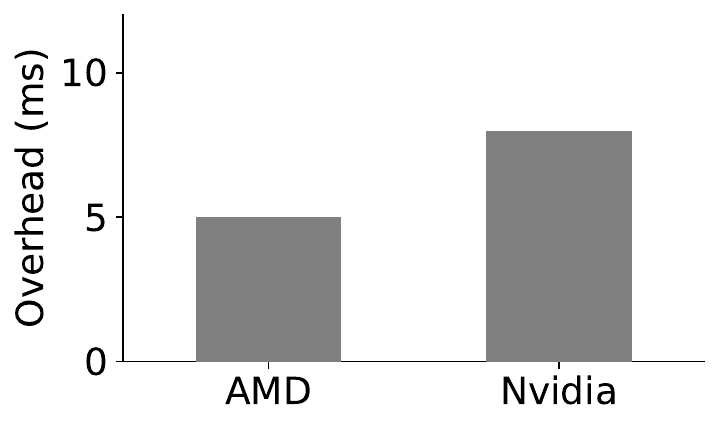}
        \caption{Predictor's overhead.\vspace{-10pt}}
        \label{fig:acc_overhead}
    \end{subfigure}
    \begin{subfigure}[!t]{0.49\linewidth}
        \includegraphics[width=\linewidth]{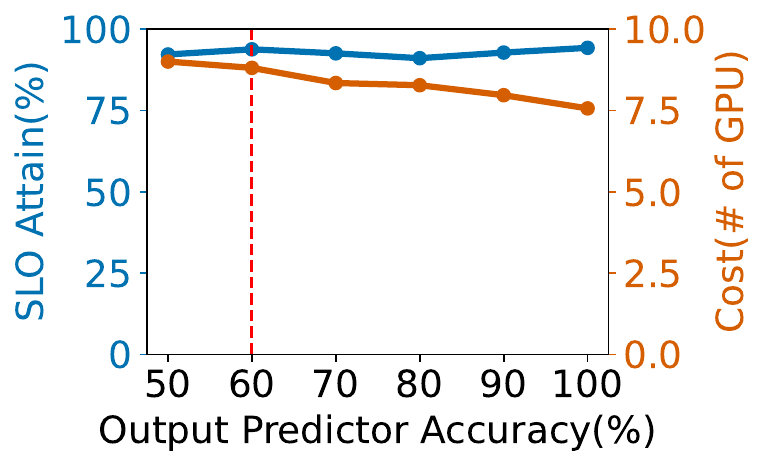}
        \caption{Sensitivity study.\vspace{-10pt}}
        \label{fig:accuracy}
    \end{subfigure}
    \caption{\sys's output predictor study includes (a)  overhead on two GPU clusters for the ShareGPT trace. (b) sensitivity of \sys to its output predictor accuracy. Red dashed line denotes measured accuracy on ShareGPT trace.}
\end{figure}
% \cz{same information as previous sentence.}

% This indicates that though our predictor cannot achieve the absolute accuracy in the prediction of the output token length (which is about $84.9\%$), it can also achieve stable end-to-end performance 
% In our following evaluation, we use an output predictor accuracy of $84.9\%$, consistent with ~\cite{hu:deepserve}.
\begin{figure}[!t]
    \centering
    \begin{minipage}[t]{0.48\linewidth}
        \centering
        \includegraphics[width=\linewidth]{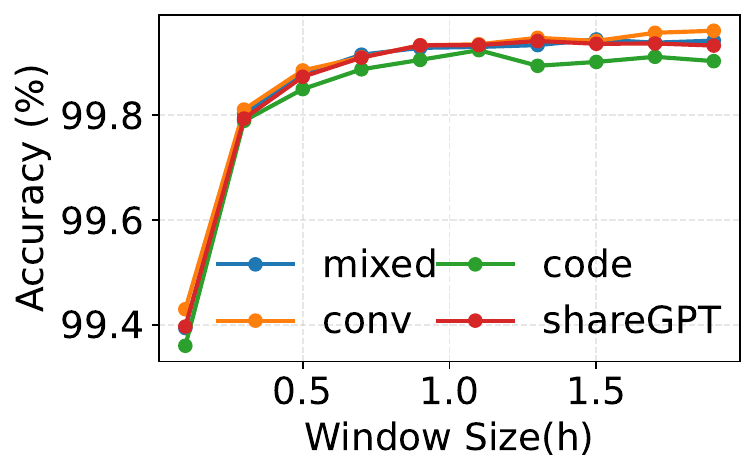}
        \caption{Online profiler accuracy vs. window size. The profiler predicts Convertible Decoder capacity for the next interval based on traffic observed in the previous window.
        }
        \label{fig:profile_accuracy_sweep}
    \end{minipage}\hfill
    \begin{minipage}[t]{0.48\linewidth}
        \centering
        \includegraphics[width=\linewidth]{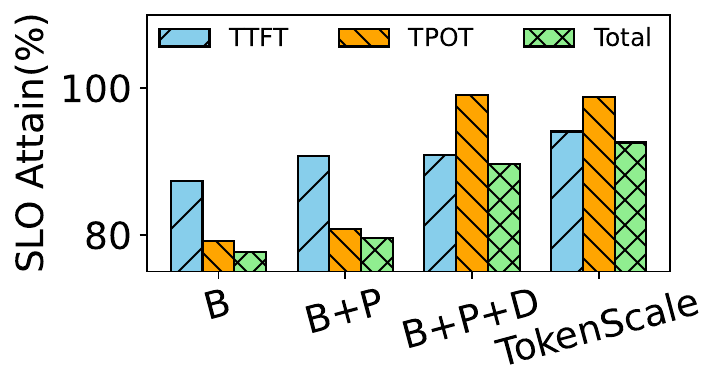}
        \caption{SLO attainment of DistServe (B), its versions with \sys's prefiller autoscaler (B+P) and \sys's prefiller and decoder autoscalers (B+P+D), and \sys.
        }
        \label{fig:ablation}
    \end{minipage}
\end{figure}
\subsubsection{Online profiler accuracy.}
\label{sec:online_profiler_acc}
% \rui{Rev A's concern} \dmi{what changed? and which concern?}
\sys's online profiler sizes the Convertible Decoder pool based on the maximum burst size observed in the previous time window. We evaluate its accuracy by sweeping the window size from 1min to 2 hours. As shown in Fig.~\ref{fig:profile_accuracy_sweep}, the accuracy approaches 99.9\% with a 1-hour window, the configuration used in our evaluation, showing that the provisioned Convertible Decoders suffice to absorb the vast majority of bursts.

\subsection{Ablation Study}
\label{sec:ablation}

We evaluate the contribution of each \sys component to SLO attainment with the mixed trace. We use \dist as the baseline (B), as it outperforms other baselines under the same cost, then replace its prefiller autoscaler with that of \sys (B+P), and further add the \sys decoder autoscaler (B+P+D), i.e., \sys without Convertible Decoders.
As shown in Fig.~\ref{fig:ablation}, the baseline achieves an overall attainment of $78\%$. B+P improves TTFT attainment from $87\%$ to $91\%$, as the \sys prefiller autoscaler scales prefillers promptly instead of waiting for queue buildup. B+P+D further boosts TPOT attainment from $80\%$ to $99\%$ and overall attainment to $90\%$, owing to the coordinated scaling of prefillers and decoders. Finally, adding Convertible Decoders raises TTFT attainment to $94\%$ without significantly affecting TPOT, as they smoothly absorb prefill surges. Overall, the full \sys design achieves the best SLO attainment.
\vspace{-10pt}
\section{Related Work}
\label{sec:related}

% \dmi{compare/contrast to splitwise is essential here. Make sure to mention that our conv decoders apply SLO-aware restricted chunk-prefill rather than be a naive fallback. }

% \textbf{Serving LLM under burst statically.}
%TODO: Discuss how other work handle the bursty workloads without autoscaling
\textbf{LLM serving without scaling.}
Many prior works handle burst workloads of LLMs without autoscaling. Some methods rely on SLO-aware routing strategies that drop requests when necessary~\cite{chen:slos-serve,zhu:cannikin} while others~\cite{duan:muxserve,kwon:efficient,zhu:cannikin,xia:skywalker-eurosys26} overprovision resources to mitigate performance degradation. Works like~\cite{li:alpaserve,su:morphserve-mlsys26} adjust model and KVC configurations to react to traffic changes. In contrast, \sys handles bursts through timely and accurate scaling, obviating the need for dropping requests, overprovisioning, and heavy modification to model structure. Related to PD-disaggregated serving, Mooncake~\cite{qin:mooncake} optimizes the KV-cache-centric serving substrate, while Beyond the Buzz~\cite{mitra:beyond-mlsys26} studies the design space of disaggregated inference. \sys instead addresses online stage-aware autoscaling for bursty PD workloads.

\noindent\textbf{Autoscaling in Conventional Serverless Systems.} Prior work in serverless autoscaling focused on request-level prediction to pre-warm function instances and mitigate cold starts~\cite{shahrad:serverless,liu:jiagu,roy:icebreaker,joosen:how,mittal:mu,cvetkovic:dirigent,singhvi:atoll}. Others optimize warm-start paths and instance caching~\cite{fuerst:iluvatar,mvondo:ofc,fuerst:faascache}. Open-source systems such as Knative~\cite{knative,k8s-hpa} adapt utilization-based metrics with panic windows for fast reaction. While valuable, these methods are ill-suited for PD-disaggregated LLM serving, as they focus on making fast scaling decisions, whereas we show that for PD systems scaling speed is much more critical than scaling decision-making due to the high initialization time of the instances.
% \rui{because new instances require substantial time to initialize, so the bottleneck lies in scaling speed rather than decision-making speed.}
In contrast, \sys is stage-aware by design, accounting for the different speeds of scaling the prefill and decode stages, and their position in the PD pipeline, and using the Convertible Decoder to bridge the initialization gap.

\noindent\textbf{Autoscaling LLM serving systems.}
To improve resource efficiency in LLM serving, many approaches explore autoscaling for LLM serving that draw inspiration from the above works.
% ~\cite{lv:dilu,wu:loongserve,du:ecoserve,miao:spotserve,sun:llumnix,yu:lambdascale,zhu:polyserve,hu:deepserve,lin:planck,choukse:dynamollm,patke:chiron,xiang:aegaeon,aibrix,zhong:distserve}. 
% \dmi{can we break the above sequence into groups? don't give them all at once with a generic comment}
% \cz{Inspired by the traditional serverless system}, 
Some systems~\cite{aibrix,zhong:distserve,zhang:blitzscale,lv:dilu,xiang:aegaeon} adopt request-based autoscaling policies, which fail to accurately capture underlying resource bottlenecks. As we show in \S\ref{sec:token_velocity},
other systems~\cite{aibrix,sun:llumnix,li:flowkv,xu:slinfer-hpca26,yu:faascale-mlsys26,zhu:polyserve} scale based on GPU utilization or end-to-end performance metrics, but these metrics lag behind request and token burst arrival. 
Recent systems~\cite{liao:dopd,lin:workload,jaiswal:sageserve} dynamically adjust the P/D ratio or instance count based on queueing feedback or workload forecasting, but such reactive or forecast-driven scaling cannot timely capture short-lived request and token bursts.
In contrast to retrofitting the policies and mechanisms from cloud computing or relying on reactive or forecast-drive capacity scaling, \sys uses the leading Token Velocity metric and the Convertible Decoder mechanism to drive system scaling, avoiding performance interference between the co-running prefill and decode tasks, as in \spw's mixed-pool PD instances~\mbox{\cite{patel:splitwise}}, by design.
% \rui{DistServe~\cite{zhong:distserve} provisions P/D capacity from offline-profiled static throughput estimates, which do not capture the runtime variation of decoder demand across workloads with different input/output-length distributions. \spw~\cite{patel:splitwise} supports bidirectional conversion between prefill and decode instances, but neither bounds the resulting P/D interference nor distinguishes the different burst characteristics of the two stages.
% Recent works also adapt PD-disaggregated or LLM serving resources dynamically. 
% Unlike these methods, which either rely on reactive or forecast-driven capacity scaling~\cite{liao:dopd,lin:workload,jaiswal:sageserve} or borrow decode capacity without bounding interference~\cite{patel:splitwise,liao:dopd}, \sys combines stage-specific provisioning with bounded borrowing to bridge the scale-out latency while limiting the interference with decode requests.}

\noindent\textbf{Accelerating LLM instance initialization.}
% Efficient autoscaling of LLMs requires rapid model initialization. 
Several works focus on reducing the initialization delays of LLM instances.
% \dmi{again, break into groups}
Some works optimize critical steps in model initialization including CUDA Graph capturing~\cite{zeng:medusa}, model weights loading~\cite{fu:serverlessllm,zhang:blitzscale,liu:pipeboost}. Other works mitigate initialization overhead by reusing inference engine components~\cite{xiang:aegaeon,hu:deepserve}.
While these approaches reduce initialization time, achieving zero-delay tolerance for prefillers remains a challenge. Moreover, \sys is complementary to these methods, which can further reduce the decoder start-up time.
% \vspace{-3pt}

\section{Conclusion}
The rise of disaggregated LLM serving architectures requires a move beyond reactive autoscaling. We presented \sys, a proactive scaling framework that combines two innovations: Token Velocity, a leading metric that drives scaling of prefill, network, and decode resources, and the Convertible Decoder to adapt to bursty workloads. Our evaluation shows that by scaling accurately and rapidly, \sys achieves lower latency and higher utilization than state-of-the-art systems under volatile workloads.
% While this work assumes scaling without prefix-caching, many production systems use multi-level, locality-aware KVC hierarchies. Token Velocity can be combined with such KVC designs to jointly address bottlenecks. Co-designing \sys with hierarchical KVC architectures is an important future direction. 

\begin{acks}
We thank the anonymous reviewers for their valuable feedback, the HyScale members for the fruitful discussions about this work, and Yao Fu for helpful feedback and writing suggestions. This research is supported by the Ministry of Education, Singapore, under its Academic Research Fund Tier 2 (MOE-T2EP20124-0002) and Tier 1 Seed (RS26/23).
\end{acks}

\newpage
\balance

% \bibliographystyle{plain}
%\bibliography{asplos25-templates/Exported_Items}
% \bibliography{Exported_Items}
\bibliographystyle{ACM-Reference-Format}
\bibliography{./bibcloud/gen-abbrev,dblp,ref}

% \newpage
% \appendix
% \input{sec/appendix}

\end{document}